\documentclass[%
reprint,
twocolumn,
superscriptaddress,
showpacs,preprintnumbers,
nofootinbib,
aps,
prd
]{revtex4-1}

\usepackage{amsmath}
\usepackage{amssymb}
\usepackage{mathtools}

\usepackage{hyperref}

\usepackage{graphicx}

\usepackage[dvips]{color}

\def\al{\alpha} 
\def\be{\beta} 
\def\ga{\gamma}
\def\de{\delta}
\def\ep{\epsilon}
\def\ze{\zeta}
\def\et{\eta}
\def\th{\theta}

\def\ka{\kappa}
\def\la{\lambda}

\def\si{\sigma}
\def\ta{\tau}

\def\Ga{\Gamma}

\def\Om{\Omega}

\newcommand{\ben}{\begin{equation}}
\newcommand{\een}{\end{equation}}
\newcommand{\bea}{\begin{eqnarray}}
\newcommand{\eea}{\end{eqnarray}}
\newcommand{\ba}{\begin{array}}
\newcommand{\ea}{\end{array}}
\newcommand{\bit}{\begin{itemize}}
\newcommand{\eit}{\end{itemize}}
\newcommand{\vev}[1]{\left\langle#1\right\rangle}

\newcommand{\mpl}{m_{\text{P}}}

\newcommand{\half}{\frac12}

\newcommand{\bk}{\textbf{k}}

\newcommand{\bx}{\textbf{x}}

\newcommand{\debar}[2]{{(2\pi)^{#1}}\de({#2})}

\newcommand{\Tfield}{\tau^\phi}
\newcommand{\Tfluid}{\tau^\text{f}}
\newcommand{\fluidT}{\tau_\text{f}}

\newcommand{\cs}{c_\text{s}} 
\newcommand{\TN}{T_\text{N}} 
\newcommand{\tN}{t_\text{N}} 
\newcommand{\Tc}{T_\text{c}} 
\newcommand{\vw}{v_\text{w}} 
\newcommand{\fieldV}{\overline{U}_\phi} 
\newcommand{\fluidV}{\overline{U}_\text{f}}  
\newcommand{\fluidVperp}{\overline{U}_\text{f}^\perp}  
\newcommand{\fluidVmax}{\overline{U}_\text{f, max}}  
\newcommand{\fluidVmaxperp}{\overline{U}_\text{f, max}^\perp}  
\newcommand{\Rb}{R} 
\newcommand{\Rc}{R_\text{c}} 
\newcommand{\Rbc}{R_*} 
\newcommand{\Hc}{H_*} 
\newcommand{\etaS}{\eta_\text{s}} 
\newcommand{\zetaB}{\ze_\text{b}} 
\newcommand{\Tvisc}{\tau_{\eta}} 
\newcommand{\Nb}{N_\text{b}}

\newcommand{\tLife}{\tau_\text{v}}
\newcommand{\Rfluid}{L_\text{f}}
\newcommand{\Rphi}{L_\phi}

\newcommand{\quadPar}{\gamma}
\newcommand{\cubPar}{A}
\newcommand{\relgamma}{W}
\newcommand{\strengthPar}[1]{\alpha_{#1}}

\newcommand{\SpecDen}[1]{P_{#1}}
\newcommand{\SpecDenGW}{\tilde P_{\text{GW}}}

\newcommand{\rGW}{\rho_\text{GW}}
\newcommand{\OmGW}{\Omega_\text{GW}}
\newcommand{\OmGWscaled}{\tilde\Omega_\text{GW}}

\newcommand{\uetcTen}{\Pi^2}

\newcommand{\IntSca}{\xi}

\newcommand{\CorLen}{\ell}

\newcommand{\phiAtMin}{\phi_\text{b}}
\newcommand{\VacEne}{V_0}

\newcommand{\Vol}{{\mathcal V}}

\definecolor{newgreen}{RGB}{10,100,20}

\begin{document}

\newcommand{\Sussex}{\affiliation{
Department of Physics and Astronomy,
University of Sussex, Falmer, Brighton BN1 9QH,
U.K.}}

\newcommand{\HIPetc}{\affiliation{
Department of Physics and Helsinki Institute of Physics,
P.O. Box 64, 
FI-00014 University of Helsinki,
Finland
}}

\newcommand{\Stavanger}{\affiliation{
Institute of Mathematics and Natural Sciences,
University of Stavanger,
4036 Stavanger,
Norway
}}

\title{Numerical simulations of acoustically generated gravitational waves at a first order phase transition}
\author{Mark Hindmarsh}
\email{m.b.hindmarsh@sussex.ac.uk}
\Sussex
\HIPetc
\author{Stephan J. Huber}
\email{s.huber@sussex.ac.uk}
\Sussex
\author{Kari Rummukainen}
\email{kari.rummukainen@helsinki.fi}
\HIPetc
\author{David J. Weir}
\email{david.weir@uis.no}
\Stavanger

\date{January 7, 2016}

\begin{abstract}
We present details of numerical simulations of the gravitational radiation produced by a first order thermal phase transition in the early universe. 
We confirm that the dominant source of gravitational waves is 
sound waves generated by the expanding bubbles of the low-temperature phase.  We demonstrate that the sound waves have a power spectrum with a power-law form 
between the scales set by the average bubble separation (which sets the length scale of the fluid flow $\Rfluid$) and the bubble wall width. The sound waves generate gravitational waves whose power spectrum also has a power-law form, 
at a rate proportional to $\Rfluid$ and the square of the fluid kinetic energy density. We identify a dimensionless parameter $\OmGWscaled$ characterising the efficiency of this ``acoustic'' gravitational wave production whose value is $8\pi\OmGWscaled \simeq 0.8 \pm 0.1$ across all our simulations.  
We compare the acoustic gravitational waves with the standard prediction from the envelope approximation. Not only is the power spectrum steeper (apart from an initial transient)
but the gravitational wave energy density is generically 
larger by the ratio of the Hubble time to the phase transition duration, which can be
2 orders of magnitude or more in a typical first order electroweak phase transition.
\end{abstract}
\preprint{HIP-2015-13/TH}
\maketitle

\section{Introduction}

Gravitational waves (GWs) promise a new and exciting window to the cosmos. Two ground-based interferometer experiments, LIGO and VIRGO,  are about to restart operations with greatly increased sensitivity \cite{Harry:2010zz,Accadia:2009zz}, and will be joined in a few years by KAGRA \cite{Somiya:2011np}.
Once working at their design sensitivity, they are expected to quickly detect gravitational wave signals from binary neutron stars \cite{Abadie:2010cf}. 
Gravitational waves also offer a unique way to learn about the early universe. A range of phenomena, such as inflation, topological defects and phase transitions may lead to observable gravitational wave signals across a wide range of frequencies (for a review see \cite{Binetruy:2012ze}). 
There are a number of proposals to realise a gravitational wave detector in space, in the first place eLISA \cite{Seoane:2013qna}, which is scheduled for launch in 2034. Space-based detectors have much longer arm lengths than ground based ones, and have maximum sensitivity in a frequency range which is relevant for a first order phase transition at the electroweak scale.

Given these exciting observational prospects, we revisit the generation of gravitational waves in first order thermal phase transitions in the early universe. 
We have in mind
an electroweak-scale phase transition, but nothing in our formalism is specific to electroweak scale physics.  In the Standard Model the electroweak transition is known to be a cross-over \cite{Kajantie:1995kf,Kajantie:1996mn,Gurtler:1997hr,Csikor:1998eu,DOnofrio:2014kta}, which does not lead to a gravitational wave signal.  However, a strong first order phase transition is possible in various extensions of the Standard Model~\cite{Carena:1996wj,Delepine:1996vn,Laine:1998qk,Grojean:2004xa,Huber:2000mg,Huber:2006wf,Dorsch:2013wja}.

We reduce the original physical system to a model consisting of a scalar order parameter field coupled to an ideal fluid. The parameters of the model can in principle be fixed by matching to the thermodynamical quantities of the original theory.  We perform very large scale numerical simulations  to determine the fluid and gravitational wave power spectra. The ultimate goal is to understand what information on the phase transition can be extracted from the future observation of a gravitational wave signal. 

Since the early nineties there have been a number of studies of gravitational waves from phase transitions. In Refs.~\cite{Turner:1990rc,Kosowsky:1991ua,Kosowsky:1992rz,Kosowsky:1992vn}, the case of a scalar field only, i.e.\ a vacuum transition without fluid, was considered, motivated by models of inflation terminated by a first order transition.  Vacuum transitions during inflation and with a fluid were considered in Ref.\ \cite{Chialva:2010jt}.

In a vacuum transition, all the energy released goes into the bubble wall, which as a result is accelerated to the speed of light.
After solving numerically the field equations for the collision of two scalar bubbles \cite{Hawking:1982ga,Kosowsky:1991ua}, it was realised that the energy-momentum tensor sourcing the gravitational wave production can be approximated by the envelope of the colliding bubbles \cite{Kosowsky:1992rz,Kosowsky:1992vn}. This ``envelope approximation'' models a configuration of expanding bubbles by the overlap of a corresponding set of infinitely thin shells. The envelope disappears once the transition is completed and gravitational wave production stops. It is found that the gravitational wave spectrum peaks at a frequency determined by the average bubble size at collision. In the UV, the spectrum falls as a power law, subsequently shown to be $k^{-1}$ \cite{Huber:2008hg}, where $k$ is the wave number. 
Numerical studies in Ref.~\cite{Child:2012qg} did not have the dynamic range to clearly confirm this behaviour, but the larger simulations done for the present work show some supporting evidence.

The case of a thermal phase transition, where the scalar field is coupled to a fluid, is more complicated. The nucleated bubbles will show accelerated expansion until the pressure inside is balanced by friction caused by the plasma. The bubbles then expand with a constant velocity. This is because the energy released by the transition grows with the volume of the bubble, i.e. $\sim R^3$, while the energy transferred to the scalar bubble wall only grows with the bubble surface, i.e. $\sim R^2$, where $R$ denotes the bubble radius. Hence only a tiny fraction of the released energy, on the order of the  ratio of the initial to final bubble radius, stays in the scalar field. In the case of a first order electroweak scale thermal phase transition this ratio is about $10^{-4}M_W/M_{\rm Pl}~\sim10^{-13}$. Therefore gravitational wave production in thermal phase transitions is completely dominated by the fluid.\footnote{An exception may be the case where the bubble wall ``runs away'', i.e. friction is not sufficient to prevent the wall from approaching the speed of light~\cite{Bodeker:2009qy}, similar to a vacuum transition. Then both the scalar and the fluid could contribute sizeably to the generated gravitational wave signal.} The energy which is released into the fluid mostly goes into reheating the plasma. A small and calculable fraction \cite{Kamionkowski:1993fg,Espinosa:2010hh} goes into bulk motion of the fluid and can source gravitational waves. 

Having established the fluid as the main source of gravitational waves, the question of the production mechanism arises. Several mechanisms have been suggested and studied in the literature. In the simplest approach, one assumes that the fluid put into motion by the scalar wall can still be treated as a thin shell and the energy momentum tensor sourcing gravitational wave production can again be approximated by the shell overlaps~\cite{Kamionkowski:1993fg}.
In this case gravitational wave production finishes with the completion of the phase transition, and a characteristic prediction is the $k^{-1}$ UV power law of the spectrum.
Another possibility is that the collision of bubbles induces turbulent motion of the fluid \cite{Kamionkowski:1993fg}. The resulting eddies generate gravitational waves even after the transition is completed~\cite{Kamionkowski:1993fg,Caprini:2006jb,Gogoberidze:2007an,Caprini:2009yp}. Various UV power laws of the the gravitational wave spectrum have been suggested in this context, such as $k^{-3.5}$~\cite{Dolgov:2002ra} and $k^{-8/3}$~\cite{Caprini:2006jb}.

To shed light onto these competing scenarios, we recently performed large scale numerical simulations of a thermal phase transition of a scalar field plus fluid system~\cite{Hindmarsh:2013xza}. We found no indications that fluid turbulence was an important source of gravitational radiation. Instead sound waves are generated by the explosive bubble growth, which propagate through the plasma until long after the transition is completed. In our simulations these sound waves are the dominant source of gravitational waves.  After the phase transition, the fluid energy-momentum tensor  clearly does not show the form assumed in the envelope approximation. The nearly linear behaviour of sound waves is very different to the highly nonlinear behaviour of the scalar field. 

Other numerical simulations of the generation of gravitational waves by the coupled field-fluid system, using an explicit update algorithm for the fluid, have been described recently in Ref.~\cite{Giblin:2014qia}. 
The generation of gravitational waves through sound in QCD and electroweak phase transitions was also recently studied in Ref.~\cite{Kalaydzhyan:2014wca}, with special focus on the effect of possible non-linear sound dispersion relations, which were argued to lead to an inverse acoustic cascade. In Ref.~\cite{Ghiglieri:2015nfa}, generation of gravitational waves in the Standard Model in the absence of such a cascade was discussed.

In the present work we simulate at larger volumes and larger average bubble separations than in \cite{Hindmarsh:2013xza}, for the same range of bubble wall speeds and phase transition strengths. We widen the dynamic range even more by nucleating all bubbles at the same time. 
We confirm that the gravitational wave density parameter is proportional to the fourth power of the mean square fluid velocity, the ratio of lifetime of the source to the Hubble time, and the ratio of length scale of the source to the Hubble length. 
We measure the length scale of the source, approximately the average bubble separation in \cite{Hindmarsh:2013xza}, directly from the fluid flow. With this improvement,  the proportionality constant for the gravitational wave density parameter varies much less between phase transitions with different strengths and bubble wall speeds. Our measurements show that it is $0.8 \pm 0.1$, where the uncertainly is the root mean square fluctuation between simulations.

We show that the resulting gravitational wave spectrum exhibits UV power laws which are clearly steeper than the $k^{-1}$ predicted by the envelope approximation. In the case of deflagrations (where the bubble walls are subsonic), we are reasonably confident that the power law is $k^{-3}$. For detonations we do not have sufficient dynamic range to be certain of the power law index. 

We compare the acoustic gravitational waves with the standard prediction from the envelope approximation. 
We argue that the envelope approximation is based on an incorrect picture of the dynamics of the fluid, in which the fluid perturbations are destroyed by bubble collisions in the same way as the bubble walls. Instead, they pass through one another, and keep oscillating, resulting in a gravitational wave source whose effective lifetime is the
Hubble time.
The true gravitational wave energy density is therefore a factor $\beta/\Hc$ higher, where $\Hc$ is the Hubble rate at the phase transition, 
and $\beta^{-1}$ is the duration of the phase transition. 
For a thermal electroweak-scale phase transition, the gravitational wave signal is larger than hitherto believed by at least two orders of magnitude.

\section{First order phase transitions in cosmology}

\subsection{Hydrodynamics}

We describe the phase transition using the cosmic fluid -- order parameter field model~\cite{Ignatius:1993qn,KurkiSuonio:1995vy}, which we summarise here.  The model contains a classical scalar field $\phi$ (effective order parameter), which is coupled to ideal fluid hydrodynamics.  The variables describing the local state of the matter are local temperature $T$, fluid 4-velocity $U^\mu$ and the scalar order parameter field $\phi$.  The first order dynamics are obtained by introducing a temperature dependent effective potential $V(\phi,T)$.  
Following \cite{Enqvist:1991xw,Ignatius:1993qn}, we use a simple $\phi^4$ form for the potential:
\begin{equation}
\label{e:ScaPot}
V(\phi, T) = \frac{1}{2} \gamma (T^2-T_0^2) \phi^2 - \frac{1}{3} \cubPar T \phi^3 + \frac{1}{4}\lambda\phi^4.
\end{equation}
The detailed form of the potential is not important, as long as it allows for a first order phase transition of sufficient strength.  A first order transition occurs if $2\cubPar^2 < 9\lambda\gamma$.

The equation of state of the coupled scalar field and fluid system is
\begin{align}
\epsilon(T,\phi) &= 3 a T^4 + V(\phi,T) - T\frac{\partial V}{\partial T},\\
  p(T,\phi) &= a T^4 - V(\phi,T)
\end{align}
where $a=(\pi^2/90)g_*$, and $g_*$ is the effective number of degrees of freedom.
The latent heat density (usually just called the latent heat) is
\ben
{\cal L}(T) = w(T,0) - w(T,\phiAtMin)
\een
where $w = \ep + p$ is the enthalpy density, and $\phiAtMin$ is the equilibrium value of the field in the symmetry-broken phase at temperature $T$.
The strength of the transition can be characterised by the ratio of the latent heat to the total radiation density in the high temperature symmetric phase,
\ben
\strengthPar{T} = \frac{{\cal L}(T)}{3aT^4}.
\label{e:StrParA}
\een
The total energy-momentum tensor of the system can be written as
\begin{equation}
\label{eq:tmunu}
T^{\mu\nu} = \partial^\mu \phi \partial^\nu \phi - {\textstyle \half} g^{\mu\nu} (\partial\phi)^2
+ \left[\epsilon + p \right] U^\mu U^\nu + g^{\mu\nu} p
\end{equation}
where the metric convention is (-+++).
The energy-momentum tensor is conserved, $\partial_\mu T^{\mu\nu} = 0$.
The interaction between field gradients and the fluid is introduced by
splitting the conserved current nonuniquely into field and fluid parts, 
which are then coupled together through a dissipative term proportional to field gradient:
\begin{align}
  [\partial_\mu T^{\mu\nu}]_{\rm field} &=
    (\partial_\mu\partial^\mu\phi) \partial^\nu\phi -
    \frac{\partial V}{\partial\phi} \partial^\nu \phi = 
    \delta^\nu 
    \label{t-munu-1}\\
    [\partial_\mu T^{\mu\nu}]_{\rm fluid} &=
   \partial_\mu[(\epsilon+p)U^\mu U^\nu] - \partial^\nu p +
    \frac{\partial V}{\partial \phi} \partial^\nu \phi = 
    -\delta^\nu,
    \label{t-munu-2}
\end{align}
where the coupling term is 
\ben
\label{e:CouTer}
\delta^\nu = \eta U^\mu \partial_\mu\phi \partial^\nu \phi
\een
with $\eta$
an adjustable friction parameter \cite{Ignatius:1993qn}. 
Equations analogous to Eqs.~(\ref{t-munu-1})-(\ref{t-munu-2}) can, at least in principle, be derived from field theory (see e.g.\ \cite{Moore:1995si,Konstandin:2014zta}), but
the simplified model here is adequate for parametrising the entropy
production~\cite{KurkiSuonio:1996rk}.

From Eqs.~(\ref{t-munu-1}) and~(\ref{t-munu-2}) we can derive the equations of
motion in a form suitable for numerical simulation.  For the field we obtain
\begin{equation}
- \ddot{\phi} + \nabla^2 \phi - \frac{\partial V}{\partial \phi} = \eta \relgamma (\dot{\phi} + V^i \partial_i \phi)\,,
\label{eqfield}
\end{equation}
where $W$ is the relativistic $\gamma$-factor and $V^i$ is the fluid 3-velocity,
$U^i = WV^i$.  For the fluid energy density $E=W\epsilon$, contracting $[\partial_\mu T^{\mu\nu}]_{\rm fluid}$ with $U_\nu$ gives
\begin{multline}
\dot{E} + \partial_i (E V^i) + p [\dot{\relgamma} + \partial_i (\relgamma V^i)]  - \frac{\partial V}{\partial \phi} \relgamma (\dot{\phi} + V^i \partial_i \phi) \\ = \eta \relgamma^2 (\dot{\phi} + V^i \partial_i \phi)^2.
\label{eqE}
\end{multline}
Finally, the equations of motion for the fluid momentum density $Z_i = W(\epsilon+p) U_i$ are
\begin{equation}
\dot{Z}_i + \partial_j(Z_i V^j) + \partial_i p + \frac{\partial V}{\partial \phi} \partial_i \phi  = -\eta \relgamma (\dot{\phi} + V^j \partial_j \phi)\partial_i \phi.
\label{eqZ}
\end{equation}
The implementation of Eqs.~(\ref{eqfield})-(\ref{eqZ}) on a discrete lattice is described in Section \ref{sect:numerical}.

The parameters of the potential in Eq.~(\ref{e:ScaPot}) are related to thermodynamic quantities at the phase transition: the critical temperature $T_c$, latent heat ${\cal L}(\Tc)$, surface tension $\sigma$ and the broken phase correlation length (which is also of order the bubble wall thickness) $\ell$~\cite{Enqvist:1991xw}
\begin{align}
  T^2_c &= \frac{T_0^2}{1-2\cubPar^2/(9\lambda\gamma)} \\
  {\cal L} &= \frac{\cubPar^2\gamma}{\lambda^2} T_0^2 T_c^2 \\
  \sigma &= \frac{2\sqrt{2}}{81}\frac{\cubPar^3}{\lambda^{5/2}} T_c^3 \\
  \ell^2 &= \frac{9\lambda}{2\cubPar^2} \frac{1}{T_c^{2}}.
  \label{e:PhaTraPar}
\end{align}
Due to supercooling, the phase transition (bubble nucleation) starts at temperature $T_N$, where $T_0 < T_N < T_c$.  We are mostly interested in the large supercooling (LSC) case, where $T_N$ is typically somewhere in the middle between $T_0$ and $T_c$.  
However, we emphasise that our focus in this work is not the nucleation of critical bubbles, which in a given microscopic theory is a thermal field theory problem and can be studied in perturbation theory 
or with numerical simulations~\cite{Moore:2000jw}.  In our simulations both the density of the initial bubbles 
and the nucleation temperature $T_N$ are set by hand.

\subsection{Bubble nucleation}
\label{ss:BubNuc}

The phase transition proceeds by the nucleation and growth of bubbles of the broken phase~\cite{Linde:1978px,Linde:1981zj}.
Bubble nucleation occurs at an exponentially growing rate per unit volume below the critical temperature~\cite{Hogan:1984hx,Enqvist:1991xw},
\ben
p(t) \simeq \Ga_0  e^{-S(\tN) + \be(t - \tN)},
\label{e:TunRat}
\een
where $-\be$ is the time derivative of the action of the critical bubble $S(t)$, and $\Ga_0$ is a dimensional prefactor of order $\alpha_W^5 \Tc^4$ \cite{Moore:2000jw}, where $\alpha_W \approx 1/30$. The nucleation time $\tN$ can be defined to be the time at which the nucleation rate reaches one bubble per Hubble volume per Hubble time, or $p(\tN) = H^4(\tN)$. 

The tunnelling rate parameter $\be$ not only sets the timescale of the transition, but also 
the average separation between bubbles once the transition has completed, $\Rbc$. 
Having defined $\Rbc$ to be the inverse cube root of the number density of bubbles, 
it can be shown that \cite{Enqvist:1991xw}
\ben
\Rbc = (8\pi)^{\frac13} \frac{\vw}{\be}.
\label{e:RbcBet}
\een
Strictly, Eq.~(\ref{e:RbcBet}) applies only for detonations.  For deflagrations, one should take into account the suppression of the tunnelling rate ahead of the bubble wall, where the fluid is heated by the release of latent heat.  In this case we would expect $\Rbc \sim  {\cs}/{\be}$.

The important ratio ${\be}/{\Hc}$ (the transition rate relative to the Hubble rate) follows from simple considerations of the temperature of the transition \cite{Hogan:1984hx}.
One can straightforwardly argue that 
\ben
S(\tN) \sim 4\ln (\mpl/\TN),
\een
and that for tunnelling in a thermal effective potential (\ref{e:ScaPot})
\ben
\frac{\be}{\Hc} \simeq \frac{2 S(\tN)}{(1 - \TN/\Tc)}.
\een
Hence, for a thermal electroweak-scale transition, the critical bubble action must be $\mathrm{O}(10^2)$, and the ratio $\be/\Hc$ must be at least $\mathrm{O}(10^2)$.

A detailed non-perturbative evaluation of the bubble nucleation rate in the standard model electroweak theory is presented in Ref.~\cite{Moore:2000jw}, using an unphysically small Higgs mass in order to ensure a first order phase transition.  In this case the critical bubble action was found to be $\approx 90$, and $\be/\Hc \approx 2\times 10^4$.
These are expected to be generic numbers for any first order thermal electroweak-scale transition.

\section{Theory of GW generation}
\label{s:TheGWGen}
\subsection{GW power spectrum definition}

A gravitational wave is a propagating mode of the transverse and traceless part of the metric perturbation, $h_{ij}$.
We are interested in calculating the gravitational wave energy density power spectrum, where the gravitational wave energy-momentum tensor is 
\begin{equation}
T_{\mu\nu}^\text{GW} = \frac{1}{32\pi G}\left< \partial_\mu h_{ij} \partial_\nu h_{ij} \right>.
\end{equation}
To this end, we define the spectral density of the time derivative of the metric perturbation $\SpecDen{\dot h}(\bk,t)$ by 
\ben
\vev{\dot{h}_{ij}(\bk,t) \dot{h}_{ij}(\bk',t) } = \SpecDen{\dot h}(\bk,t) \debar3{\bk+\bk'}.
\een
The gravitational wave energy density power spectrum is then
\ben
\frac{d \rGW}{d \ln(k)} = \frac{1}{32\pi G} \frac{k^3}{2\pi^2} \SpecDen{\dot h}(\bk,t).
\label{e:GWPowSpeDef}
\een

\subsection{GW power spectrum from fluid and field}

The source of gravitational waves is the transverse traceless part of the spatial components of the energy-momentum tensor.  
Given that we will be removing the trace anyway, it suffices to consider a source tensor
$\tau_{ij}  = \Tfield_{ij} + \Tfluid_{ij}$, which is decomposed into fluid and field pieces according to 
\ben
\label{e:TauDef}
\Tfield_{ij} = \partial_i \phi \partial_j \phi, \quad\Tfluid_{ij} = \relgamma^2 (\epsilon + p)V_i V_j.
\een
The physical metric perturbations are recovered in momentum space by 
applying the projector onto transverse, traceless symmetric rank 2 tensors:
\ben
\label{e:ProDef}
\la_{ij,kl}(\bk) = P_{ik}(\bk) P_{jl}(\bk) - \half P_{ij}(\bk)P_{kl}(\bk) 
\een
with 
\ben
P_{ij}(\bk)  = \de_{ij} - \hat{k}_i \hat{k}_j. 
\een
The particular solution for the gravitational wave is therefore
\ben
h_{ij} (\mathbf{k},t) = (16\pi G)\la_{ij,kl}(\bk) \int_0^t dt' \frac{\sin[k(t-t')]}{k} \ta_{kl}(\bk,t'),
\label{e:GWsol}
\een
where we have assumed that the source vanishes for $t'<0$.

Using the fact that the fluid shear stress dominates the spatial parts of the energy-momentum tensor, we write 
\begin{widetext}
\bea
\vev{\dot{h}^{ij}_\bk(t) \dot{h}^{ij}_{\bk'}(t) } &=&  (16\pi G)^2 \int^t_0 dt_1 dt_2 \cos[ k(t - t_1)] \cos[ k(t - t_2)] 
\la_{ij,kl}(\bk)\vev{\fluidT^{ij}(\bk,t_1)\fluidT^{kl}(\bk',t_2)}.
\eea
Introducing the unequal time correlator (UETC) of the fluid shear stress $\uetcTen$ \cite{Caprini:2009fx,Figueroa:2012kw} through 
\ben
\la_{ij,kl}(\bk)\vev{\fluidT^{ij}(\bk,t_1)\fluidT^{kl}(\bk',t_2)} = \uetcTen(k,t_1,t_2) \debar3{\bk + \bk'}
\een
and averaging over a period $T$, much longer than the periods of the gravitational waves of interest, we can 
write
\ben
\SpecDen{\dot h}(k,t) = (16\pi G)^2 \int^t_0 dt_1 dt_2 \frac{\cos[ k(t_1 - t_2)]}{2} \uetcTen(k,t_1,t_2).
\een
\end{widetext}
On dimensional grounds, we can write the UETC as 
\ben
\label{e:UETCmod}
\Pi^2(k,t_1,t_2) \simeq [(\bar\ep+\bar p)\fluidV^2]^2\Rfluid^3\tilde\Pi^2
\een
where 
$\bar\ep$ and $\bar{p}$ are the spatially averaged energy density and pressure;
$\fluidV$ is the root mean square fluid velocity, 
defined through
\bea
\label{e:fluidVdef}
(\bar\ep+\bar p) \fluidV^2 &=&  \frac{1}{\Vol}\int_{\Vol} d^3x\Tfluid_{ii}, 
\eea
where $\Vol$ is the averaging volume;
$\Rfluid$ is a characteristic length scale in the velocity field; and $\tilde\Pi^2$ is a dimensionless function of $k$, $t_1$ and $t_2$.
In Ref.~\cite{Hindmarsh:2013xza} we estimated that $\Rfluid$ would be the mean bubble separation, but we will not make that assumption yet. We will see that we can understand the numerical results better if we extract the scale directly from the fluid velocity field in the simulations.

We also assumed that the UETC would be a function of $t_1 - t_2$ for times between the nucleation time $\tN$ and the lifetime of the velocity perturbations $\tLife$, and that there is no separate timescale in the function $\tilde\Pi^2$, apart from that generated from $\Rfluid$, the speed of sound $\cs$, and the speed of light.
With these assumptions, we can write the dimensionless UETC as a function of $k\Rfluid$ and $z = k(t_1 - t_2)$, 
and the spectral density of $\dot h$ becomes
\begin{widetext}
\ben
\SpecDen{\dot h}(k,t) = \left[16\pi G(\bar\ep+\bar p)\fluidV^2\right]^2  t k^{-1} \Rfluid^3 \int dz \frac{\cos(z)}{2} \tilde\Pi^2(k\Rfluid,z).
\label{e:SpeDenExpA}
\een
Note that one could follow through the same arguments for the scalar field, which would contribute in exactly an analogous manner
\ben
\SpecDen{\dot h}^\phi(k,t) = \left[16\pi G(\bar\ep+\bar p)\fieldV^2\right]^2  t k^{-1} \Rphi^3\int dz \frac{\cos(z)}{2} \tilde\Pi^2_\phi(k\Rphi,z),
\een
\end{widetext}
where 
\bea
\label{e:fieldVdef}
(\bar\ep+\bar p) \fieldV^2 &=&  \frac{1}{\Vol}\int_{\Vol} d^3x\Tfield_{ii}, 
\eea
$\Rphi$ is a characteristic scale in the scalar field configuration, and $ \tilde\Pi^2_\phi$ is the dimensionless unequal time correlator of the scalar field shear stress tensor.   However, as explained in the introduction, the field contribution is negligible in most phase transitions.

Hence, putting together (\ref{e:GWPowSpeDef}) and (\ref{e:SpeDenExpA}), we may write
the gravitational wave energy density power spectrum as 
\ben
\frac{d \rGW}{d \ln(k)} = 8 \pi G\left[(\bar\ep+\bar p)\fluidV^2\right]^2  t \Rfluid \frac{(k\Rfluid)^3}{2\pi^2} \SpecDenGW(k\Rfluid),
\label{e:GWEneDenPowSpe}
\een
where
\ben
\SpecDenGW(k\Rfluid)  = \frac{1}{k\Rfluid}\int dz \frac{\cos(z)}{2} \tilde\Pi^2(k\Rfluid,z),
\label{e:NoDimSpecDen}
\een
is a dimensionless spectral density for the gravitational waves. 

The gravitational wave power spectrum at time $t$ can then be written
\bea
\rGW &=& (\bar\ep+\bar p)^2 \fluidV^4 (t\Rfluid) (8\pi G \OmGWscaled),
\label{e:RhoGWdot}
\eea
where 
\ben
\OmGWscaled = \int \frac{dk}{k}\frac{(k\Rfluid)^3}{2\pi^2} \SpecDenGW(k\Rfluid)
\een
is a dimensionless number.  
We see that the gravitational wave power spectrum grows linearly with time, for as long as the velocity perturbations are active, with a slope which depends on the square of the enthalpy density, the fourth power of the mean square fluid velocity, the fluid length scale, and a dimensionless number describing the fluid flow $\OmGWscaled$.

In principle, the value of $\OmGWscaled$ depends on the parameters of the phase transition in dimensionless combinations, which we can expect to include the bubble wall speed $\vw$ and the latent heat relative to the total energy density $\strengthPar{T}$.  
In Fig.\ 2 (bottom) of \cite{Hindmarsh:2013xza}, we plotted $\rGW/\left[(\bar\ep+\bar p)^2 \fluidV^4 \Rfluid\right]$ against time. 
Noting that we have $G=1$, the slope of the graph is $8\pi \OmGWscaled$. 
We found that $\OmGWscaled$ was approximately constant, varying by no more than a factor 2, when 
we took the fluid scale $\Rfluid$ to be the mean bubble separation $\Rbc = \sqrt[3]{V/\Nb}$. 
Hence most of the dependence of the gravitational radiation energy density on the phase transition parameters is accounted for by the explicit factors in Eq.~(\ref{e:RhoGWdot}).

\subsection{Integral scale}
\label{ss:IntSca}

The question of which scale to take for $\Rfluid$ affects the value of $\OmGWscaled$, and hence its variation between simulations.  
As mentioned above, in Ref.~\cite{Hindmarsh:2013xza} we took the scale to be $\Rbc$, the average bubble separation.
However, one could equally estimate the length scale from the velocity field itself, and to this end we can use the following quantity  (sometimes referred to as the integral scale) 
\ben
\IntSca_\text{f} =   \frac{1}{\vev{V^2}}\int \frac{d^3k}{(2\pi)^3} |k|^{-1} \SpecDen{V}(k),
\een
where $\vev{V^2}$ is the RMS velocity.
We will see that when the scale $\Rfluid$ is chosen to be the integral scale, the variation in the parameter $\OmGWscaled$ is reduced to about 10\%.  
This emergence of $\OmGWscaled$ as a quasi-universal constant for first order phase transitions with $\strengthPar{\TN} \lesssim 0.1$ is an important result.

One can also define an integral scale $\IntSca_\text{GW}$ for the gravitational wave energy density from its spectral density $\SpecDenGW$.  We will also confirm that the integral scale of the gravitational radiation is related to the integral scale of the velocity field, as one would expect.

\begin{widetext}
\subsection{Dimensionless GW power spectrum parameter $\OmGWscaled$ }

It is often useful to express the gravitational wave power spectrum as a fraction of the critical density, $\rho_\text{c} = 3H^2/8\pi G$. Hence we are led to consider 
a dimensionless gravitational wave power spectrum 
\ben
\frac{d \OmGW(k,t)}{d \ln(k)} =  \left[16\pi G(\bar\ep+\bar p)\fluidV^2\right]^2   \frac{t \Rfluid}{\Hc^2} \frac{(k\Rfluid)^3}{24\pi^2} \SpecDenGW(k\Rfluid),
\een
where $\Hc$ is the Hubble parameter at the time the bubbles are nucleated. Noting that the critical density is the energy density $\bar\ep$, and denoting the lifetime of the source by $\tLife$, we find that the dimensionless gravitational wave power spectrum during the radiation era can be written   
\bea
\label{e:GWPowSpe}
\frac{d \OmGW(k)}{d \ln(k)} &=&  3(1+w)^2 \fluidV^4  (\Hc \tLife) (\Hc\Rfluid) \frac{(k\Rfluid)^3}{2\pi^2} \SpecDenGW(k\Rfluid),
\eea
\end{widetext}
where $w = \bar{p}/\bar{\ep}$ is the equation of state parameter.
Integrating over wavenumber, we see that the total relative energy density is
\ben
\OmGW = 3(1+w)^2 \fluidV^4 (\Hc \tLife) (\Hc\Rfluid) \OmGWscaled.
\label{e:OmgwEqn}
\een

\subsection{Source lifetime}
\label{ss:LifSouWav}

It is clearly important for the calculation of the gravitational wave energy density to calculate the lifetime of the source, the shear stress caused by the sound waves. 
We show in Appendix \ref{s:GWExpUni} that in an expanding universe, the shear stresses decay and decorrelate in such a way to make $\tLife$ precisely equal to the Hubble time.
The shear stresses also decay due to the viscosity of the fluid at a scale-dependent rate. We should therefore estimate on which scales viscous damping time is smaller than $\tLife$.

For linear non-relativistic flows induced by sound waves (i.e.\ for velocity fields $V^i_\parallel$ which are purely longitudinal), viscosity adds a term of the form 
\ben
\left(\frac{4}{3}\etaS + \zetaB\right) \nabla^2 V^i_\parallel
\een
to the left hand side of Eq.\ (\ref{eqZ}), where $\etaS$ is the shear viscosity and $\zetaB$ is the bulk viscosity.  For a plasma of relativistic particles in a gauge theory, the bulk viscosity is negligible compared to the shear viscosity~\cite{Arnold:2006fz}, and the shear viscosity can be estimated as  
\ben
\etaS \sim T^3/e^4 \ln(1/e),
\een
where $e$ is the electromagnetic gauge coupling~\cite{Arnold:2000dr}.
Hence velocity perturbations of wavenumber $k$ are damped as $\exp(-4 \etaS k^2 t/3)$, and the lifetime due to viscous damping of sound waves with wavelength $R$ is  
\ben
\Tvisc(R) \sim R^2\ep/\etaS \sim e^4 \ln(1/e)R^2 T.
\een
Therefore, at a transition with temperature just below the critical temperature $\Tc$, the viscous damping lifetime exceeds the Hubble time $\Hc^{-1}$ for all scales 
\ben
R \gg \frac{\vw}{\Hc}\left(\frac{\sqrt{a}\Tc}{\mpl e^4}\right) \sim 10^{-11} \frac{\vw}{\Hc}\left(\frac{\Tc}{100\; \text{GeV}}\right),
\een
where we have neglected the logarithm of the gauge coupling.

We will see in the next section that the scale of the fluid perturbations is set by the average separation of the nucleating bubbles $\Rbc$, and that the bubble separation at an electroweak-scale phase transition with any interesting degree of supercooling will satisfy this inequality. 

Hence for a first order transition at the electroweak scale -- or even a few orders of magnitude above -- the lifetime of the 
source of the gravitational waves
 is the Hubble time,
\ben
\tLife = \Hc^{-1} \ll \Tvisc(\Rbc).
\een

\subsection{Comparison to envelope approximation}
\label{ss:ComEnvApp}
In the envelope approximation, the relative energy density in gravitational waves is given by~\cite{Kamionkowski:1993fg,Huber:2008hg}
\begin{equation}
\label{e:EnvAppFor}
\Om_\text{GW}^\text{ea} \simeq \frac{0.11 \vw^3}{0.42+\vw^2} \left( \frac{\Hc}{\be}\right)^2 \frac{\ka^2 \al^2}{(\al+1)^2},
\end{equation}
where $\al$ is the ratio between the ``vacuum'' energy (defined below) and the radiation energy density in the symmetric phase, $\kappa$ is the efficiency with which vacuum energy is converted to kinetic energy, and $\be$ is the nucleation rate parameter also defined above.

The vacuum energy $\VacEne$ is defined in Ref.~\cite{Kamionkowski:1993fg} from the trace anomaly, 
\ben
\th = \ep - 3p,
\een 
as a quarter of the difference between the symmetric and broken phases: 
\ben
\VacEne = \frac14(\th_\text{s} - \th_\text{b}).
\een
In our convention, the trace anomaly vanishes in the symmetric phase, and in the broken phase is
\ben
\label{e:VacEne}
\th_{\text{b}} =  -T \frac{d}{dT}V(\phiAtMin,T) + 4V(\phiAtMin,T),
\een
where $\phiAtMin$ is the value of $\phi$ in equilibrium in the broken phase at temperature $T$. 
In the conventions of \cite{Kamionkowski:1993fg}, the trace anomaly vanishes in the broken phase, and in the symmetric phase is equal and opposite to (\ref{e:VacEne}). 
Hence for our thermal potential (\ref{e:ScaPot}) the parameter $\al$ is 
\ben
\al = \frac{\VacEne}{3aT^4} = \frac{1}{3aT^4}\left( \frac{1}{4}T \frac{d}{dT}V(\phiAtMin,T) - V(\phiAtMin,T) \right).
\label{e:StrParB}
\een
The efficiency parameter is defined from the average fluid kinetic energy density (\ref{e:fluidVdef}) as 
\ben
\ka = \frac{1}{\VacEne} \frac{1}{\Vol} \int d^3 x \Tfluid_{ii}.
\een
Therefore
\ben
\label{e:KappaEquiv}
(1+w) \fluidV^2 = \frac{\ka\al}{1+\al}. 
\een
The factor of $1+\al$ in the denominator of the right hand side comes from the fact that we are dividing by the average total energy density in the symmetric phase, which is 
$3aT^4 + \VacEne$ in the conventions of \cite{Kamionkowski:1993fg}. 
 
Note that $\ka\al$ is conventionally estimated analytically from the radial fluid velocity around an isolated expanding bubble
$v(r,t)$, where $r$ is the distance from the centre of the bubble, and $t$ is the time since nucleation~\cite{Kamionkowski:1993fg,Espinosa:2010hh}.
At large times, the radial fluid velocity is a function of a scaling variable $\xi = r/t$, rather than $r$ and $t$ separately. 
The ratio of the kinetic energy density to the total energy density can then be estimated as  
\ben
\ka\al = \frac{3}{\vw^3 \ep} \int d\xi \xi^2 (\ep + p) \relgamma^2 v^2(\xi).
\een
We will compare this estimate to the numerically obtained $(1+w) \fluidV^2$ in the results section, finding good agreement.
 
In order to compare our expression for the gravitational wave energy density (\ref{e:OmgwEqn}) with the envelope approximation formula (\ref{e:EnvAppFor}), we estimate the fluid flow scale $\Rfluid$ as the bubble separation scale $\Rbc$, which in turn is related to the nucleation rate parameter by Eq.\ (\ref{e:RbcBet}).

Hence the ratio between the gravitational wave energy density generated acoustically and in the envelope approximation is\footnote{Note that for a deflagration, if tunnelling is suppressed behind the shock wave, the ratio is boosted by a factor $\sim \cs/\vw$ -- see the discussion after Eq.\ (\ref{e:RbcBet}).}
\ben
\frac{\OmGW}{\OmGW^\text{ea}} \simeq \frac{3(8\pi)^{\frac13}\OmGWscaled}{0.11\vw^2(0.42+\vw^2)}\left({\be}\tLife\right).
\label{e:OmGWRat}
\een
Given that the ratio (\ref{e:OmGWRat}) is smallest for $\vw=1$, and that the lifetime of the sound waves is approximately $\Hc^{-1}$ (see Section \ref{ss:LifSouWav}), we can estimate that 
\ben
\frac{\OmGW}{\OmGW^\text{ea}} \gtrsim 60 \OmGWscaled \frac{\be}{\Hc}.
\een
We will see from our numerical simulations that $\OmGWscaled \sim 0.04$.
The ratio ${\be}/{\Hc}$ was discussed in Section~\ref{ss:BubNuc}, and shown to be at least $\mathrm{O}(10^2)$, and possibly significantly greater if there is only small supercooling.
We conclude that the energy density in acoustically generated gravitational waves is at least two orders of magnitude greater than the envelope approximation suggests.

\section{Numerical simulations}
\label{sect:numerical}

\subsection{Methods}

Our numerical methods are a development of those first used in this context to study the case of isolated bubbles in Ref.~\cite{KurkiSuonio:1995vy}. In that paper a spherically symmetric bubble was assumed. Here we extend those simulations to a full 3+1-dimensional simulation volume. In addition, we couple the linearised stress-energy tensor to perturbations around a flat metric, to measure the gravitational wave power produced by the simulation.

\subsubsection{Coupled field-fluid system}

\begin{figure}[t]
\begin{centering}
\includegraphics{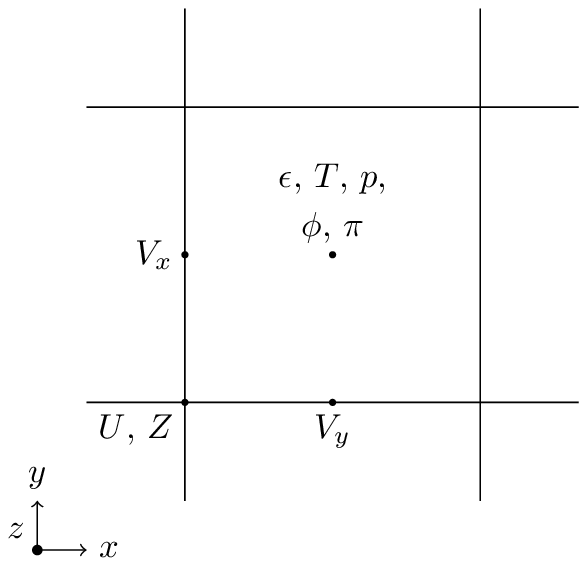}\\
\vspace{0.5cm}
\includegraphics{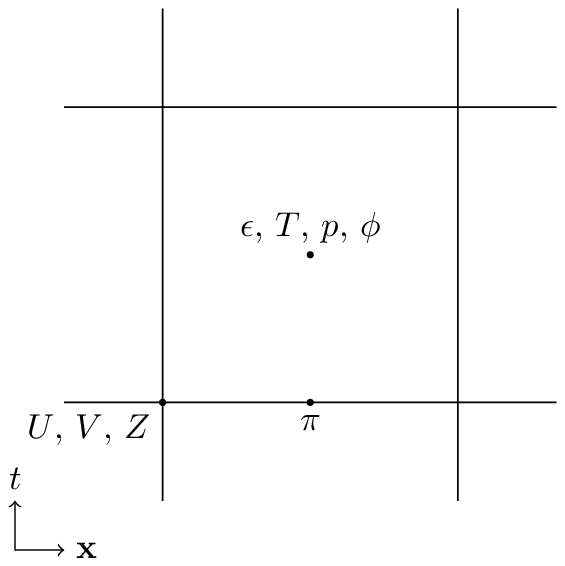}\\
\caption{\label{fig:layout} Layout of quantities simulated. The positions of quantities related to simulating an ideal relativistic fluid are standard~\cite{WilsonMatthews}. Because the field and fluid are coupled together, it is important that the scalar field $\phi$ and its conjugate momentum $\pi$ are correctly centred. We take $\phi$ to reside in zones (like pressure, temperature, etc.), so that no centring is required to compute, for example, the equation of state.}
\end{centering}
\end{figure}

The coupled hydro-scalar equations, outlined above, can be treated quite easily using standard numerical techniques. The scalar field is evolved with the leapfrog (Verlet) algorithm, while standard operator splitting methods are used for the fluid~\cite{WilsonMatthews}. These are equivalent to numerically integrating the equations of motion given above.

Although the full details of how to implement relativistic hydrodynamics is beyond the scope of this paper, it is instructive to consider how the quantities are laid out on the lattice both in the spatial and temporal directions (see Fig.~\ref{fig:layout}). Furthermore, for good energy conservation it is essential that the discretised version of the damping term couple the field and fluid quantities at equal times during the simulation.

We have tested the results of our simulations against changing timestep (as well as the lattice spacing); see the following section.

As our simulations do not run for sufficiently long to develop strong shocks (indeed, we choose our lattice spacing parameters such that the fluid velocity profile is always resolved by several $\delta x$), the simulations presented in this paper do not involve any artificial viscosity. The importance of an artificial viscosity term was previously studied using 1+1-dimensional simulations of two colliding bubbles.

\subsubsection{Metric perturbations}

Our principal observables are the energy density and power spectrum of the gravitational waves. 
The goal of our simulations is to compute  the power per unit logarithmic frequency interval in gravitational waves $d\rho_\text{GW}(k)/d\ln k$, 
and the total energy density $\rGW$.

Perturbations of the metric are sourced by transverse-traceless part of the stress-energy tensor $\Pi_{ij}$
\begin{equation}
\ddot{h}_{ij} - \nabla^2 h_{ij} = 16 \pi G \Pi_{ij}.
\end{equation}
Obtaining $\Pi_{ij}$ from $T_{\mu\nu}$ involves a projection in momentum space. Therefore, evolving $h_{ij}$ (whether in momentum space or position space)
would involve Fourier transforms at each timestep.
As we go to large volumes, the execution time of fast Fourier transforms (FFTs) scales as $\mathrm{O}(N \log N)$, while there are few optimised FFT codes that offer domain decomposition in more than one dimension. 
It is therefore vital that the number of steps requiring Fourier transforms be minimised, to yield a scalable simulation.

Our approach is to evolve the unprojected equation of motion in real space~\cite{GarciaBellido:2007af}
\begin{equation}
\label{eq:unprojected}
\ddot{u}_{ij} - \nabla^2 u_{ij} = 16 \pi G (\Tfield_{ij} + \Tfluid_{ij}),
\end{equation}
where $u_{ij}$ is an auxiliary tensor and 
the sources are defined in Eq.~(\ref{e:TauDef}).
Only when we wish to recover the metric perturbations $h_{ij}$ do we Fourier transform $u_{ij}$ and project out the transverse-traceless components through
\begin{equation}
h_{ij} (\mathbf{k}) = \lambda_{ij,lm} (\hat{\mathbf{k}}) u_{lm} (t,\mathbf{k}),
\end{equation}
where
the projector is defined in Eq.~(\ref{e:ProDef}).
We evolve Eq.~(\ref{eq:unprojected}) using a leapfrog algorithm in a similar manner to the scalar field.

Note that we choose the units of the code such that the critical temperature $\Tc = 1$ and the gravitational constant $G=1$.

\subsection{Tests}

Our basic tests principally involve varying the lattice spacing and timestep independently, on simulations of a single bubble colliding with itself in a small periodic box. These allow us to test that the simulations perform accurately between length scales $1/\Rbc$ and $1/\ell$. Longer distances do not need to be tested, and in any case $\Rbc$ is set by the box size $L$ in these tests.

\subsubsection{Changing the lattice spacing}

We performed tests on the effect of changing the lattice spacing using the self-collision of a single bubble in a cubic box, in relatively modest volumes (with parameters given in the following section). We considered $\delta x = 0.5/\Tc$, $\delta x = 1/\Tc$, $\delta x = 2/\Tc$ and $\delta x = 4/\Tc$. We notice no significant difference between these choices until $\delta x = 4/\Tc$.

\begin{figure}[t]
\begin{centering}
\includegraphics[width=0.4\textwidth,clip=true]{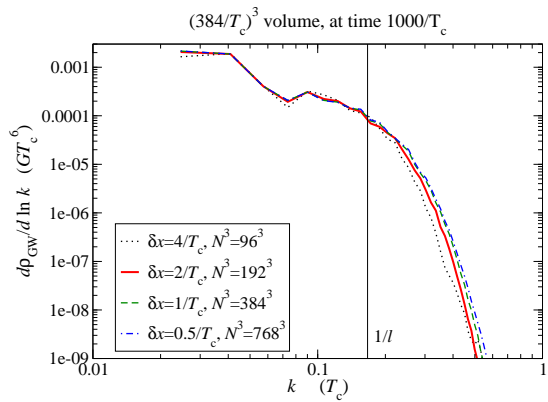}\\
\caption{\label{fig:singlebubble} Single bubble test simulation, with the correlation length $\ell$ shown as an indication of the wall width. Only the fluid source is shown here; discretisation errors for the field source are the same or smaller. There is good agreement between $1/L$ and $1/\ell$, as desired.}
\end{centering}
\end{figure}

It is worth mentioning that, even for an isolated bubble which would (in continuum) have vanishing quadrupole moment and hence not source gravitational waves, the lattice discretisation breaks the spherical symmetry and results in a small amount of gravitational wave production. 
This power goes to zero as $(\delta x)^4$ for both the field and the fluid sources. After collision, however, agreement is very good with relative differences of at most $7\%$ for $k \lesssim 1/\ell$ between $\delta x=1/\Tc$ and $\delta x=2/\Tc$ for the fluid source. Furthermore, at higher momenta there are only $\mathrm{O}(1)$ differences between these two choices, consistent across seven orders of magnitude. This is surprisingly good given the relatively coarse wall width and the complicated microphysics. Similarly, discrepancies between $\delta x = 0.5/\Tc$ and $\delta x = 1/\Tc$ at late times were at worst $2 \%$ for $k \lesssim 1/\ell$; see Figure~\ref{fig:singlebubble}. Discretisation errors were always less severe for the field source than for the fluid source.

In summary, we note no significant sensitivity to lattice spacing so long as it is kept well below the scalar field wall width. 

While our previous work used simulations with $\delta x = 1/\Tc$, we use a lattice spacing of $\delta x = 2/\Tc$ in the present paper. The inferred discrepancies are demonstrably smaller than $10\%$, and the doubling of the accessible dynamic range that this allows is very useful.

\subsubsection{Changing the timestep}

With $\delta x=2/\Tc$ having been chosen, we varied the timestep to explore the effect of inaccuracies in our evolution algorithm. There is agreement at the $1\%$ level or better for all $k \lesssim 1/\ell$ and $5\%$ or better up to $k \sim 0.5$ (all points plotted on Fig.~\ref{fig:singlebubble}) as we varied $\delta t$ between $0.2/\Tc$, $0.1/\Tc$ and $0.05/\Tc$, in the same single-bubble tests for the fluid power spectrum described above. We use $\delta t=0.1$ for the remainder of the paper, although we could probably have achieved acceptable results with $\delta t=0.2$.

In the present paper our simulation durations are typically the same order of magnitude as one light-crossing time, and rather less than one sound-crossing time. This means, in particular, that the production of gravitational radiation by acoustic waves (or by scalar radiation, which is in any case heavily damped) is not likely to be affected by signals propagating around-the-lattice.

\subsection{Parameter choices}

We use the same parameters for the potential as in our previous paper. The exact values of these parameters are not particularly important: it is the latent heat and the wall velocity which mainly determine the gravitational wave power spectrum. Our aim in the present paper is to develop the ideas underlying our previous paper as well as the spherical studies carried out earlier, and so we work with the same parameters as before.

No attempt is made in the present paper to look at strong fluid flows or fast `runaway' bubble walls. We leave these harder topics for future work and instead seek to comprehensively explain the generation of gravitational waves by more gentle phase transitions ($\alpha_{T_\mathrm{N}} \lesssim 0.1$).

We discuss bubble nucleation further in the next section but note that we nucleate all of our bubbles simultaneously in the present work.

Given $\delta x =2/\Tc$ and a simulation size of $2400^3$ points, our physical simulation volume is $(4800/\Tc)^3$ for all the results presented in this paper.

\begin{table}

\begin{tabular}{l | c | c | c |}
 & Weak & Weak (scaled)  & Intermediate \\
\hline
$T_0/\Tc$ & $1/\sqrt{2}$ & $1/\sqrt{2}$ & $1/\sqrt{2}$ \\
$\quadPar$ & $1/18$ & $4/18$ & $2/18$ \\
$\cubPar$ & $\sqrt{10}/72$ & $\sqrt{10}/9$ & $\sqrt{10}/72$ \\
$\la$ & $10/648$ & $160/648$ & $5/648$ \\
${\cal L}/T_c^4$      & $9/40$ & $9/40$ & $9/5$ \\
$\sigma/T_c^3$ & $1/10$ & $1/20$ & $4\sqrt{2}/10$ \\
$\CorLen \Tc$ & $6$ & $3$ & $6/\sqrt{2}$ \\
$\TN/\Tc$ & $0.86$ & $0.86$ & $0.8$\\
$\al_{\TN}$ & $0.010$ & $0.010$ & $0.084$ \\
$\al$ & $0.0046$ & $0.0046$ & $0.050$ \\
$\Rc \Tc$ & $ 16$ & $8.1$ & $8.6$ \\
\hline

\end{tabular}

\caption{Scalar potential parameters (\ref{e:ScaPot}), nucleation temperature $\TN$, 
phase transition parameters (\ref{e:PhaTraPar}), transition strength parameters 
(\ref{e:StrParA}) and (\ref{e:StrParB}), and critical bubble radii (\ref{eq:critbub})  for our simulations.}

\label{t:SimParsPot} 
\end{table}

\subsection{Initial conditions}

\label{sec:initconds}

At the start of our simulation, we nucleate a controllable number of bubbles, which was usually $\mathrm{O}(1000)$ (yielding bubbles of average collision radius slightly larger than in Ref.~\cite{Hindmarsh:2013xza}), but was as small as 37 or in one case as large as 32558. These have a Gaussian scalar field profile. This profile is initially at rest, meaning that the conjugate momentum, and also the fluid velocity are zero in the vicinity of the bubble. We ensure that all the initially nucleated bubbles are well separated at the start of the simulation. For runs with the same number of bubbles but different wall velocities, all bubbles are nucleated at the same positions, but from testing we found that even 37 bubbles was enough to remove any discernible dependence on the initial bubble positions.

The critical bubble radius can be computed from the surface tension $\sigma$ and the difference in potential energy at $\TN$ from the thin-wall formula (noting that the potential energy in the symmetric phase is zero)
\begin{equation}
\label{eq:critbub}
\Rc = \frac{2\sigma}{- V(\phi_b,\TN)}.
\end{equation}
Values of $\Rc$ for our simulations are shown in Table~\ref{t:SimParsPot}. Rather than find the critical bubble profile exactly, we use a spherically symmetric Gaussian field profile
\begin{equation}
\phi(r) = \phiAtMin \exp(-r^2/2\Rc).
\end{equation}
This is rather broad, and therefore sufficiently large compared to the true critical bubble profile to ensure that the bubbles reliably expand despite lattice effects.

The bubbles are sufficiently large that they immediately start growing, driven by the pressure difference between the interior and the exterior.  The  scalar field quickly settles into a kink-like configuration, interpolating between the metastable and stable minima over a distance of order $\ell$, the correlation length of the scalar field (see Table~\ref{t:SimParsPot} for the values the correlation length takes).  For the scalar field dynamics to be valid we must have a lattice spacing that resolves the wall width (see previous section), which places an upper limit on the physical simulation volume possible for a given amount of computer memory.

In this paper, the bubbles are nucleated simultaneously. 
Nucleating at a single time helps to ensure clear scale separation in the limited dynamic range available to our numerical simulations, although it does produce oscillatory patterns in the resulting power spectrum (we cover the case of unequal nucleation times in Appendix~\ref{a:BubNucTim}). We could in principle recover the power spectrum produced by bubbles of all different sizes by a linear 
superposition of the resulting power spectra, weighted by the bubble size distribution.

\begin{figure}
\begin{centering}
\includegraphics[width=0.4\textwidth,clip=true]{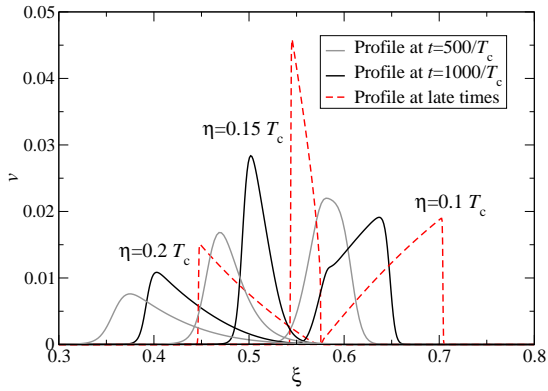}

\caption{\label{fig:profilecomparison} Comparison of radial fluid velocity profiles for simulations at approximate collision times for $\Nb=1000$ ($t=500/\Tc$; gray), $\Nb=37$ ($t=1000/\Tc$; black) and at late times (dashed red). }

\end{centering}
\end{figure}

Once nucleated, the bubbles grow, and the fluid approaches a characteristic radial velocity distribution, which is a function of $\xi=r/t$, where $r$ is the distance from the centre of the bubble, and $t$ is the time since nucleation (see Fig.~\ref{fig:profilecomparison}). The form of this function depends  on the bubble wall velocity~\cite{KurkiSuonio:1995vy}, and we will refer to it as the scaling profile. The rate of approach to the fluid scaling profile is generally much slower than the relaxation of the scalar field. 
In a true electroweak phase transition the bubble size at collision is many orders of magnitude larger than the bubble size at nucleation, giving a lot of time for the radial velocity distribution to reach its asymptotic profile. 

In our numerical simulations, the ratio of the bubble size at collision, $\Rbc$ to the bubble size at nucleation ($\approx \Rc$) is at most 90 for $\Nb=37$, $\ell=16$ and as small as 9.4 for $\Nb=32558$, $\ell=16$ (our simulation parameters are outlined in Tables \ref{t:SimParsPot} and \ref{t:SimParsRuns}).
We should therefore be alert to the fact that the fluid has definitely not settled down to its final scaling profile in a collision.  One can test for the effect of a non-scaling fluid profile by repeating simulations with fewer bubbles, so that there is a longer time before collision. We have carried out simulations such that $\Rb$ varies by around a factor of three in the two sets of simulations for which we present plots, and by as much as a factor of ten in our full set of simulations for this paper.

\begin{table}
\begin{tabular}{cccrcccc}
    Type      & $\eta/T_c$ & $\vw$ & $\Nb$ & $\fluidVmax$ & $\fluidVmaxperp$ & $\IntSca_\text{f, end} T_c$ & $8\pi\OmGWscaled$ \\
\hline
  Weak        &  0.06  &  0.83 &   988 &      0.0052  &         0.00037  &     351              &     0.88       \\
              &        &       &   125 &      0.0052  &         0.00028  &     649              &     0.84       \\
\cline{2-8}
              &  0.1   &  0.68 &   988 &      0.0084  &         0.00036  &     244              &     0.73       \\
              &        &       &   125 &      0.0082  &         0.00026  &     451              &     0.71       \\
              &        &       &    37 &      0.0080  &         0.00021  &     644              &     0.60       \\
\cline{2-8}
              &  0.121 &  0.59 &   988 &      0.0116  &         0.00052  &     182              &     0.69       \\
\cline{2-8}
              &  0.15  &  0.54 &   988 &      0.0102  &         0.00037  &     230              &     0.54       \\
              &        &       &    37 &      0.0120  &         0.00025  &     428              &     0.80       \\
\cline{2-8}
              &  0.2   &  0.44 & 32558 &      0.0059  &         0.00047  &     136              &     0.97       \\
              &        &       &   988 &      0.0073  &         0.00031  &     368              &     0.70       \\
              &        &       &   125 &      0.0075  &         0.00023  &     613              &     0.86       \\
              &        &       &    37 &      0.0078  &         0.00019  &     942              &     0.70       \\
\cline{2-8}
              &  0.4   &  0.24 &   988 &      0.0036  &         0.00049  &     756              &     0.86       \\
\hline
Wk. (sc.)     &  0.4   &  0.44 &   988 &      0.0075  &         0.00029  &     365              &     0.81       \\
\hline
Interm.       &  0.4   &  0.44 &   988 &      0.0595  &         0.00328  &     485              &     1.04       \\
\hline
\end{tabular}
\caption{\label{t:SimParsRuns} Simulation parameters $\eta$
  (field-fluid coupling), $\Nb$ (number of bubbles nucleated), with
  the resulting bubble wall speed $\vw$, the maximum fluid RMS
  velocity $\fluidVmax$, the maximum contribution of transverse fluid
  motion $\fluidVmaxperp$, the integral scale of the fluid
  $\IntSca_\text{f, end}$, and the scaled slope parameter for the
  growth of the gravitational wave energy density $\OmGWscaled$. The
  potential parameters and derived quantities for each type --
  ``weak'', ``weak scaled'' and ``intermediate'' -- are given in
  Table~\ref{t:SimParsPot}.}
\end{table}

\subsection{Scaling}

A cosmological first order phase transition is a multiscale problem, with length scales varying from the microscopic ($1/T$, bubble wall thickness) up to the Hubble scale, a range spanning 
17 orders of magnitude at the electroweak scale.
The typical bubble sizes  at collision time are somewhere between these scales, depending on the metastability of the high temperature phase.  It is of course impossible to include all of these scales in a single numerical simulation, where scale hierarchies only of order $10^2$ are achievable.  To obtain a stable numerical description of the bubble wall, the wall thickness has to span a few lattice units (denoted by $\delta x$).  In order to have collisions within the simulation volume, this restricts the bubble separation to be unphysically small.

However, this restriction can be relaxed, at least partly:
we expect that the dynamics of the bubble growth, collisions and the subsequent generation of gravitational waves are mostly determined by the ``bulk'' thermodynamics ($\epsilon$, $p$, latent heat ${\cal L}$) and the friction parameter $\eta$, but not by microscopic details of the bubble wall (surface tension $\sigma$, wall thickness $\ell$).  Dimensionally, it is clear that the contribution from quantities proportional to bubble volume (e.g. latent heat) will dominate over quantities proportional to the area of the bubbles when the bubble radius is large enough.

This motivates us to search for a way to modify the equations of motion (\ref{eqfield})--(\ref{eqZ}) so that we could simulate bubbles which are significantly larger than the microscopic length scale, while preserving the bulk thermodynamics of bubble expansion while possibly sacrificing the properties of the bubble wall.  Indeed, this can be achieved with the following simple rescaling of the parameters and fields:
\begin{equation}
  \begin{array}{l}
  \gamma \rightarrow r^2 \gamma, ~~
  \cubPar \rightarrow r^3 \cubPar, ~~
  \lambda \rightarrow r^4 \lambda, ~~
  \eta \rightarrow r \eta,  \\
  \phi(x) \rightarrow r^{-1}\phi(rx), ~~
   V^i(x) \rightarrow V^i(rx),\\
   E(x) \rightarrow E(rx),~~~~~
   Z_i(x) \rightarrow Z_i(rx),
  \end{array}
  \label{scaling}
\end{equation}
Here $r$ is a dimensionless scaling factor, and $x=(\bx,t)$.
Clearly, the equations of motion (\ref{eqfield})--(\ref{eqZ}) remain valid.  The crucial feature of the scaling is that the potential remains constant, $V(\phi,T) = V_{\rm scaled}(r^{-1}\phi,T)$, indicating that the bulk quantities $T_c$, ${\cal L}$ and also $\epsilon(T,\phi)$ and $p(T,\phi)$ remain invariant, as desired.  However, the surface tension and wall thickness scale as $\sigma\rightarrow r^{-1}\sigma$ and $\ell\rightarrow r^{-1}\ell$.  In effect the scaling stretches the field configuration by a factor of $r^{-1}$ in spatial and temporal directions.

We note that in spite of the non-trivial scaling of the friction parameter $\eta$, the total frictional force imparted on the moving bubble wall does not change: it is obtained by integrating the $\eta$-terms in Eqs.~(\ref{eqfield})--(\ref{eqZ}) over the bubble wall thickness, which is scaled by a factor of $r^{-1}$.

What does the rescaling gain us?   It is straightforward to see that the lattice implementation of the equations of motion (\ref{eqfield})--(\ref{eqZ}) do not change (in lattice units) under scaling (\ref{scaling}), provided that the lattice spacing is also scaled as $\delta x \rightarrow r^{-1} \, \delta x$.  This implies that a single lattice simulation exactly corresponds to a whole family of results, given by the scaling with $r$.  All of them have the same bulk thermodynamical properties.  Thus, provided that the detailed bubble wall properties are not important for bubble collisions and gravitational wave generation, we can take a simulation run where bubbles have been nucleated at specific locations, and rescale it to 
the desired physical bubble separation scale.

We can test the assumption that the surface properties are not important by comparing results from simulations which differ only in surface tension and wall thickness.  This can be achieved by applying the scaling (\ref{scaling}) to the parameters of the theory, but leaving the lattice spacing constant.  A set of parameters for unscaled ($r=1$) and scaled ($r=2$) runs are shown in Table \ref{t:SimParsPot}.  The surface tension and the bubble wall thickness have been halved in the scaled simulation.  The results of the test are shown in Table \ref{t:SimParsRuns}; here the scaled run is done with $\eta/\Tc=0.4$ and 988 bubbles, which can be compared with unscaled $\eta/\Tc=0.2$, 988 bubble results.  In both simulations the bubbles are nucleated at identical times and locations.  The numerical results match well within uncertainties of the measurements, supporting our assumption that the surface properties of the scalar field profile are unimportant.

\section{Numerical results}
\label{s:NumRes}

\begin{figure*}[t]

\begin{centering}
\includegraphics[height=0.29\textwidth]{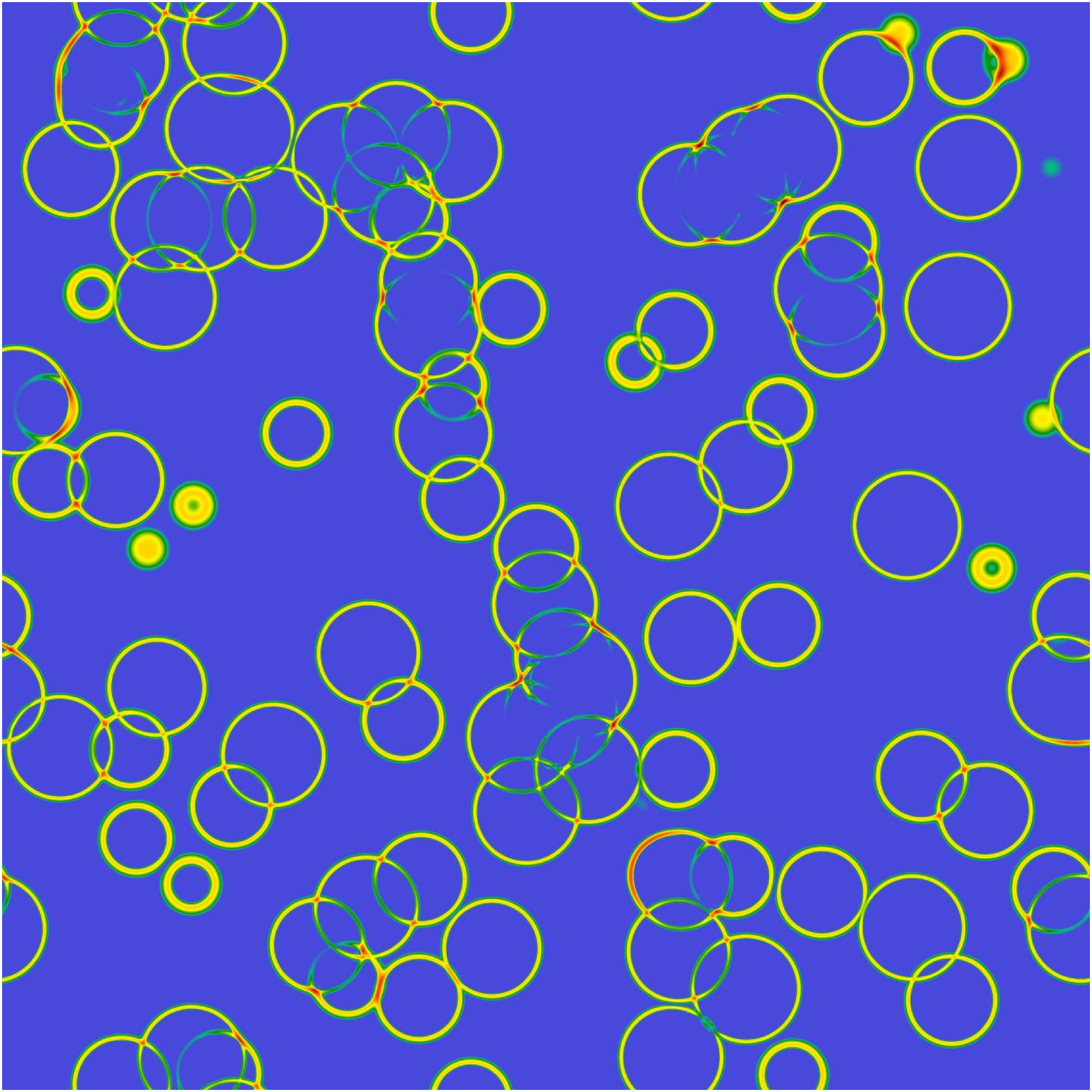} \hfill
\includegraphics[height=0.29\textwidth]{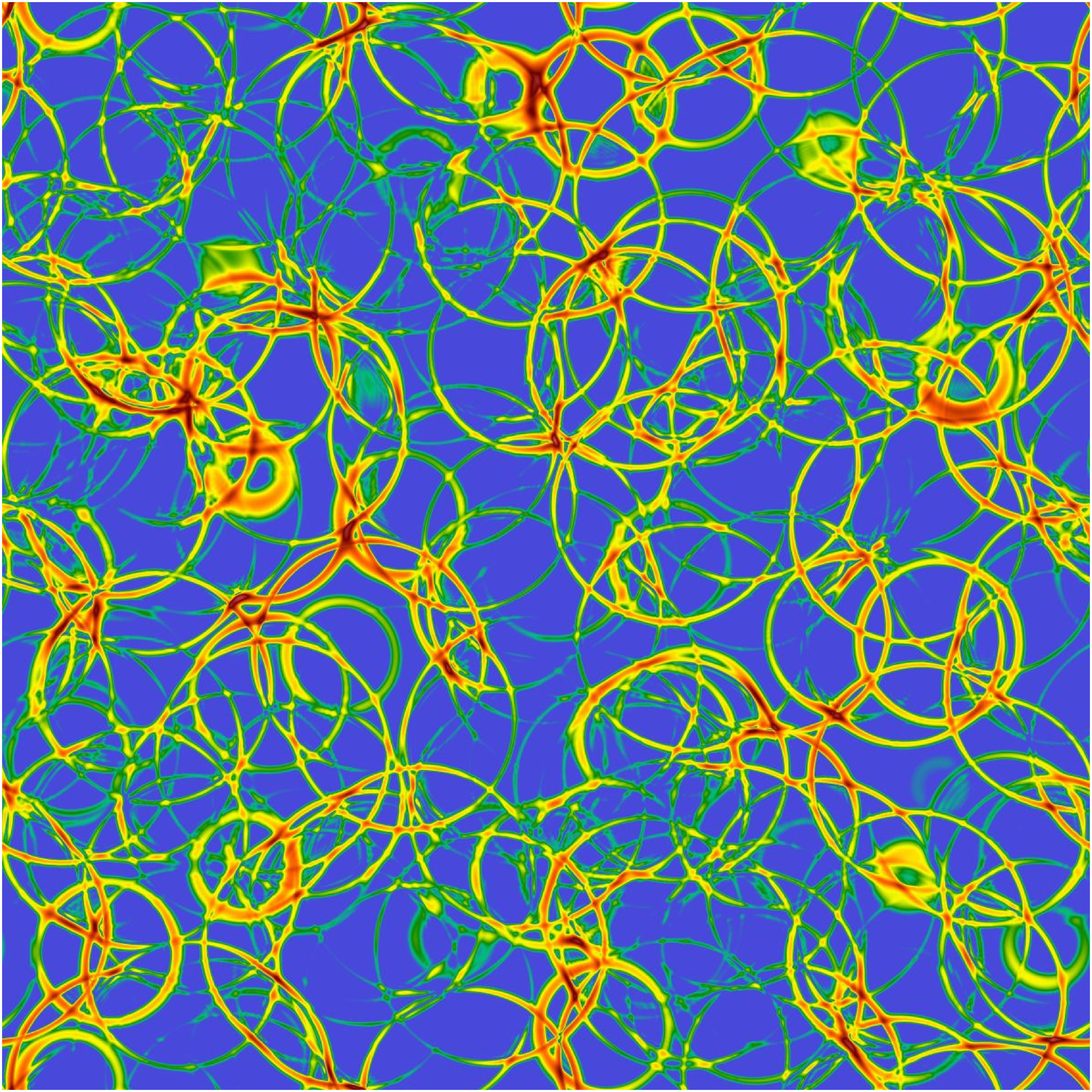} \hfill
\includegraphics[height=0.29\textwidth]{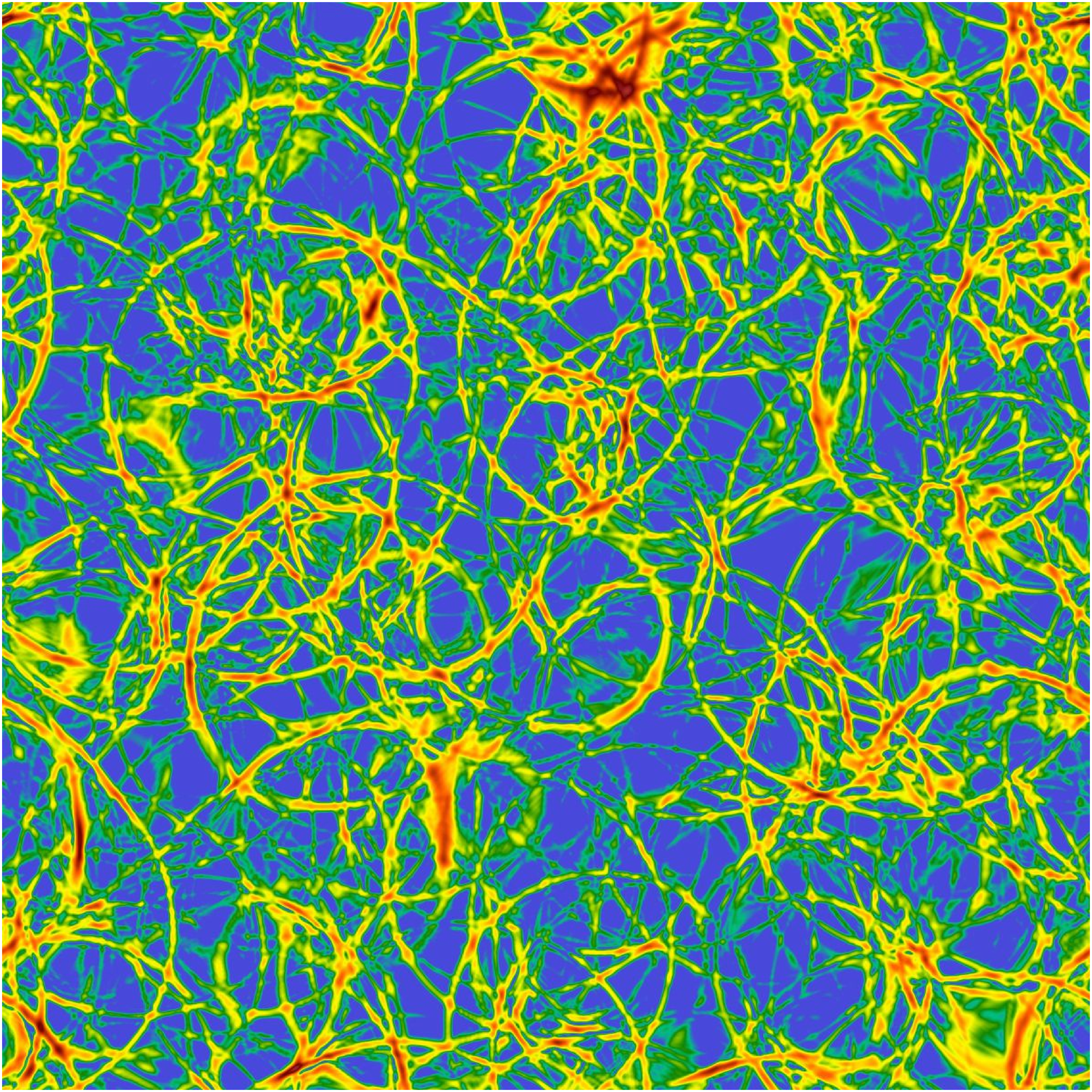} \hfill
\includegraphics[height=0.29\textwidth]{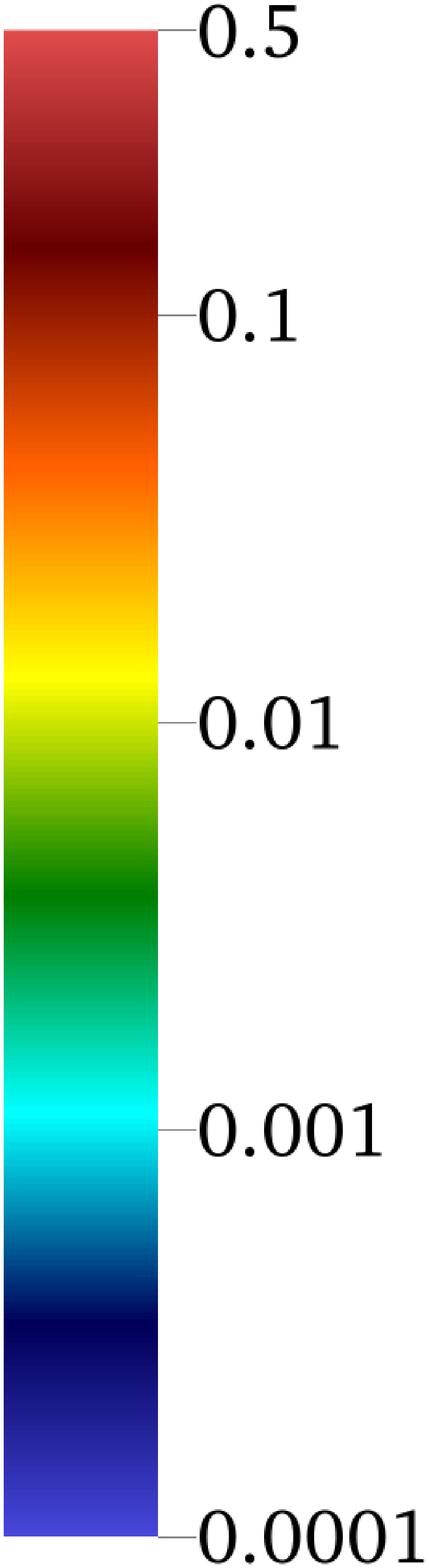}
\caption{\label{fig:slice} Slices of fluid kinetic energy density $E/\Tc^4$ at $t=500\, T_\mathrm{c}^{-1}$, $t=1000 \,T_\mathrm{c}^{-1}$ and $t=1500\, T_\mathrm{c}^{-1}$ respectively, for the $\eta/T_c=0.15$, $\Nb=988$ simulation.}
\end{centering}
\end{figure*}

In this section we present the main results from a campaign of numerical simulations, whose parameters are given in Table \ref{t:SimParsRuns}.  
As mentioned before, our simulations were carried out in a volume $(4800/\Tc)^3$.
A set of representative slices through the simulation are shown in Fig.~\ref{fig:slice}.
Our main results are derived from the set of simulations with latent heat to thermal energy ratio $\al_{\TN}\simeq 0.01$, which we characterise as a ``weak'' transition.  
Values of the friction parameter $\eta$ were chosen to give bubble growth proceeding by both detonation and deflagration, as well as one simulation tuned to the Jouguet case, where the bubble wall moves at the speed of sound. We have one ``intermediate'' strength transition, with $\strengthPar{\TN} \simeq 0.1$ where $\eta$ is chosen to give the same wall speed as the weak transition with $\eta/\Tc = 0.2$.  The ``weak scaled'' transition (employing the scaling of the previous section) is discussed later.

When plotting graphs, we focus on three representative cases, where the field-fluid coupling is $\eta/T_c=0.1$, $\eta/T_c=0.15$, and $\eta/T_c=0.2$,   
and the bubble wall speed is supersonic ($\vw = 0.83$), just subsonic ($\vw = 0.54$), and subsonic ($\vw = 0.44$). 
A complete set of graphs can be found in the supplementary material~\cite{supplementary}.

Our understanding of the transition developed in Section \ref{s:TheGWGen} shows that the important quantities for the overall gravitational wave energy density are the RMS fluid velocity $\fluidV$ and the fluid velocity scale $\Rfluid$, and 
that the gravitational wave power spectrum is only indirectly dependent on the strength of the transition and the parameters in the potential.  Indeed, the gravitational wave power spectrum should be the same for parameters which give the same $\al_{\TN}\simeq 0.01$ (keeping the wall velocity constant). We can use the ``weak scaled'' run of Table \ref{t:SimParsRuns} to test this statement, where $\al_{\TN}$ is constant but the scalar bubble wall thickness is halved.  We test the effect of the strength of the transition with the ``intermediate'' run of Table \ref{t:SimParsRuns}.

We track the progress of the transition through  
the time evolution of the two quantities $\fieldV$ and $\fluidV$ defined in Eqs.\ (\ref{e:fluidVdef}),~(\ref{e:fieldVdef}).
We recall that the squares of these quantities give an estimate of the size of the shear stresses of the field and the fluid relative to the background fluid enthalpy density, and that the $\fluidV$ tends to the RMS\ fluid velocity for $\fluidV \ll 1$. We also note that the fraction of the fluid velocity power coming from rotational modes, $\fluidVperp$, is very small leading us to conclude that rotational fluid modes are not important in this system; we discuss this in more detail in Appendix~\ref{a:TraVelNeg}.

\begin{figure}
\begin{centering}
\includegraphics[width=0.4\textwidth,clip=true]{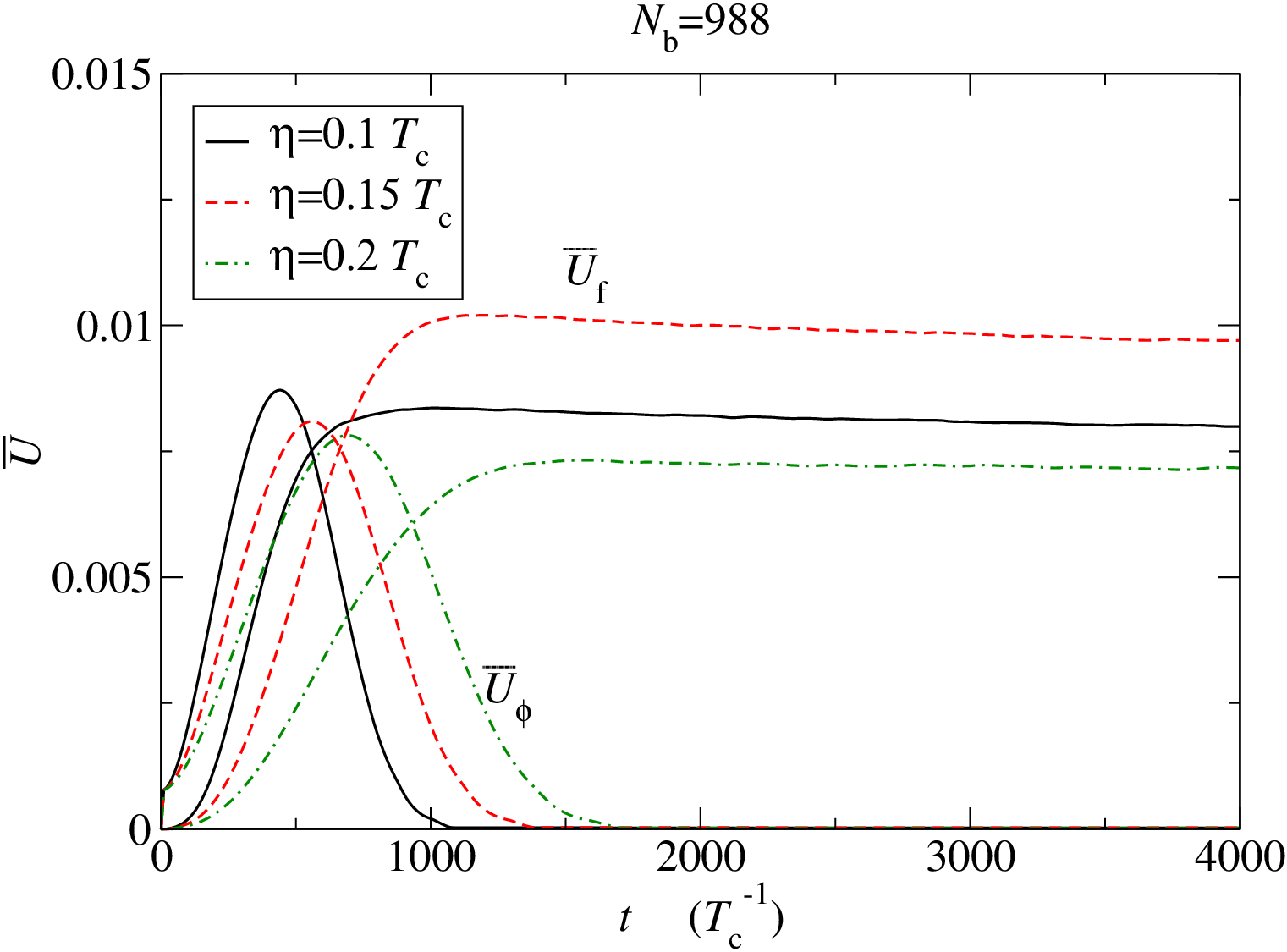}
\includegraphics[width=0.4\textwidth,clip=true]{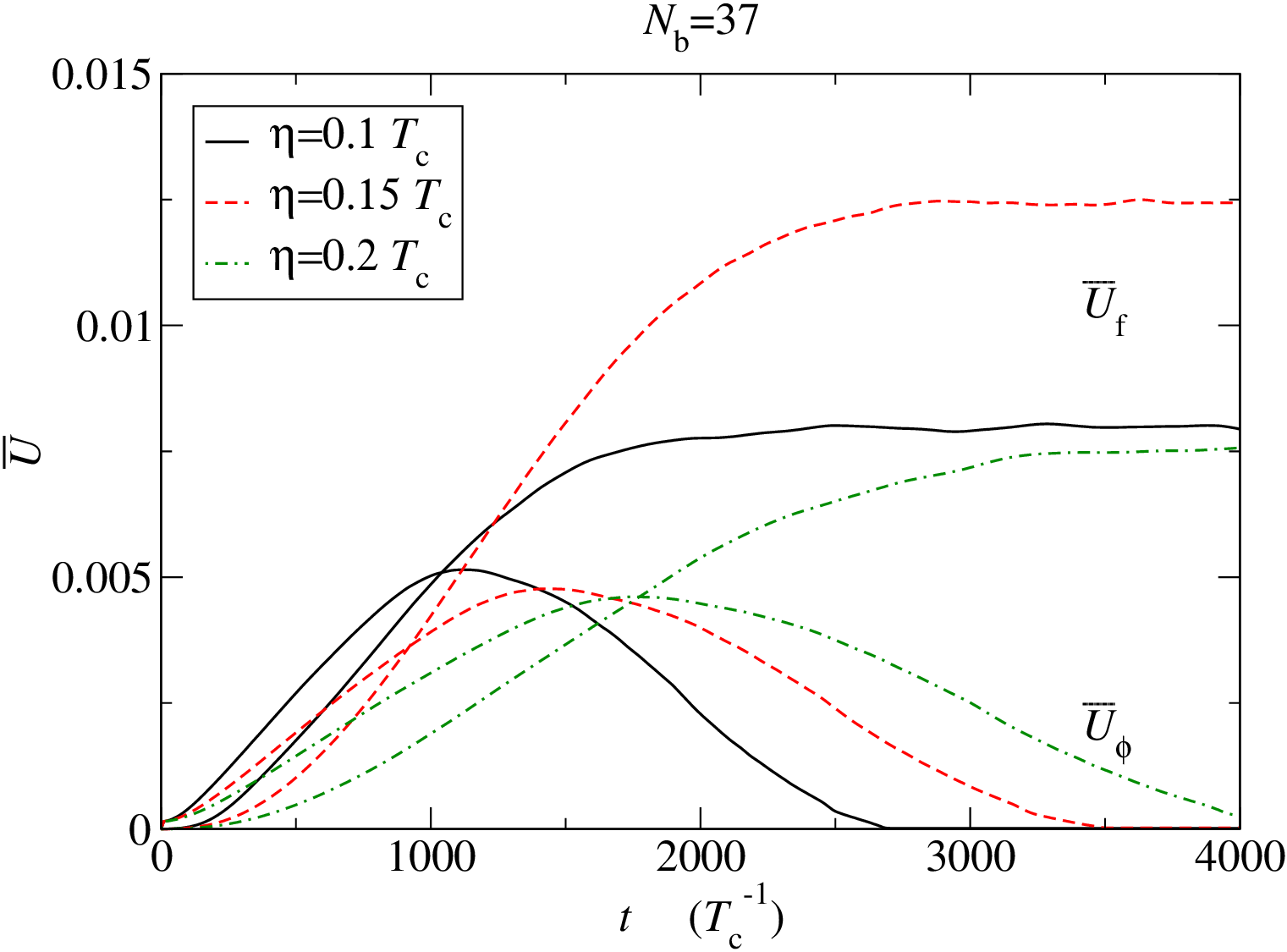}

\caption{Root mean square fluid velocity $\fluidV$ and root mean square scalar gradients $\fieldV$ for $\Nb = 988$  (top row) and $\Nb = 37$   (bottom row). }
\label{fig:timevolV} 
\end{centering}
\end{figure}

We see from Fig.~\ref{fig:timevolV} that $\fieldV$ grows and decays with the total surface area of the bubbles of the new phase, while the mean fluid velocity grows with the volume of the bubbles, and then stays constant once the bubbles have merged\footnote{We have no explicit viscosity, and the slight decreasing trend in some measurements of $\fluidV$ arises from the well-known numerical viscosity of donor-cell advection, $\nu_\text{num} \simeq \fluidV \delta x$.}.
This allows us to identify distinct phases of the transition: the collision phase, where $\fieldV$ grows and decays; and the subsequent acoustic phase where $\fluidV$ is approximately constant, and $\fieldV$ vanishes.

\subsection{Length scales}

\begin{figure}[t]
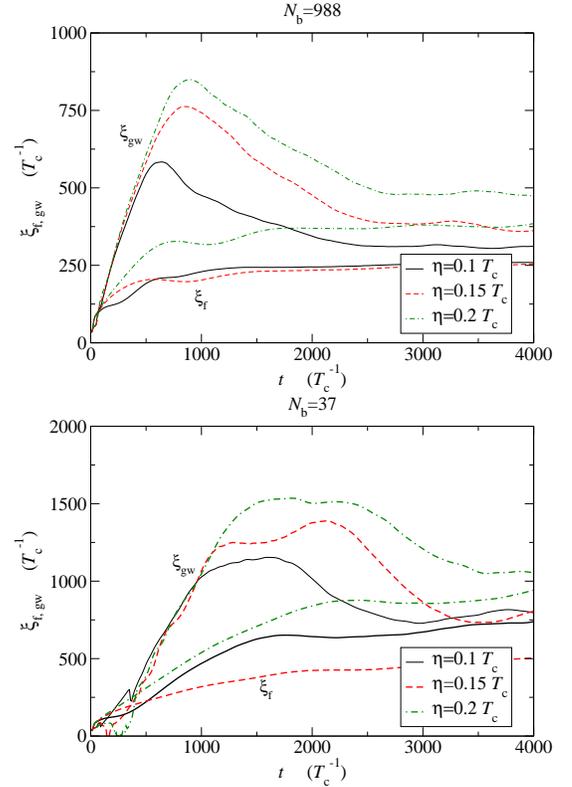

\begin{centering}
\includegraphics[width=0.4\textwidth,clip=true]{integralscale-1000.eps}
\includegraphics[width=0.4\textwidth,clip=true]{integralscale-37.eps}
\caption{\label{fig:integralscale} Plot of integral scales $\IntSca_\text{f}$ and $\IntSca_\text{gw}$ associated with the fluid and gravitational wave power spectrum for simulations of interest for $\Nb = 988$ (top) and $\Nb = 37$ (bottom). Note the different $y$-axis plotting scales.}
\end{centering}
\end{figure}

The analysis of Section \ref{s:TheGWGen} shows that the length scale of the velocity flow is an important determinant of the gravitational wave power spectrum. In Fig.\ \ref{fig:integralscale} we show the integral scales for the velocity and the gravitational radiation for runs at $\Nb=37$ and $\Nb=988$. During the collision phase, the bubbles expand and overlap, and hence the scales of the velocity field and the resulting gravitational radiation grow linearly in time. The gravitational radiation length scale is 2-3 times that of the velocity field. The scale of the velocity field stops growing as the bubbles collide and the scalar field decays to the vacuum, and stays constant during the acoustic phase. The scale imprinted on the gravitational radiation during the acoustic phase is close to that of the velocity field.

\subsection{Velocity profile}

Given the discussions on initial bubble sizes in Section~\ref{sec:initconds}, it is important to bear in mind that the bubbles in our simulations expand in size by a factor of only around 10-100,
which is many orders of magnitude less than in a real phase transition.  One practical effect is that the profile of the velocity field around the bubbles does not reach its asymptotic scaling form, which can be expressed in terms of the previously introduced ratio $\xi = r/t$. In Fig.\ \ref{fig:profilecomparison} we showed the velocity profiles for the weak transition at $\eta/T_c=0.1$, $\eta/T_c=0.15$, and $\eta/T_c=0.2$, after times $t = 500/\Tc$ and $t = 1000/\Tc$.  These are approximately when most bubble collisions are happening, in the $\Nb = 1000$ and $\Nb = 37$ runs respectively.

We see that, at collision, the velocity profiles are qualitatively similar to their asymptotic forms in amplitude and shape, but differ in detail. In particular, the peak velocities are lower. 
This is particularly noticeable at the earlier time. We would therefore expect the RMS velocities $\fluidV$ measured in the simulations to be underestimates.  As the gravitational wave power spectrum depends on the fourth power of $\fluidV$, this is a significant source of uncertainty in deriving accurate predictions for the gravitational wave power spectrum.

\begin{table}[h!]
\begin{tabular}{ccccc}
     $\eta/T_c$ & $\vw$ &  $\fluidV$ &  $\sqrt{\frac{3}{4}\ka_\text{1d}\al}$  &  $\sqrt{\frac{3}{4}\ka_\text{Esp}\al}$   \\
\hline
                0.06  &  0.83 &        0.0052     &      0.0056      &      0.0063      \\
                0.1   &  0.68 &         0.0084     &       0.0085     &       0.0121      \\
                0.121 &  0.59 &         0.0116     &      0.0146     &      0.0192     \\
                0.15  &  0.54 &         0.0102     &       0.0103     &   0.0100      \\
                  0.2    & 0.44      &        0.0073     &   0.0066     &      0.0065        \\
                0.4   &  0.24 &         0.0036     &       0.0033     &      0.0036     \\
\hline
\end{tabular}
\caption{\label{tab:tablethree} Simulation parameters $\eta$ (field-fluid coupling), with the resulting bubble wall speed $\vw$, fluid RMS velocity $\fluidV$, for weak transitions with $\Nb=988$, and the equivalent quantity $\sqrt{4\ka\al/3}$ appearing in the envelope approximation (see Eq.\ \ref{e:KappaEquiv}). The efficiency parameter $\ka$ is estimated in two ways: 
$\ka_\text{1d}\al$ is estimated from the numerical spherically symmetric 1D fluid profiles at $t=1000/\Tc$, while 
$\ka_\text{Esp}$ comes from the function 
$\ka(\vw,\alpha)$ given in the Appendix of Ref.~\cite{Espinosa:2010hh}, using 
$\vw$ extracted from spherical 1D simulations at $t=1000/\Tc$.
   }
\end{table}

These considerations are tested in Table \ref{tab:tablethree}, where we compare our mean square fluid velocity parameter $\fluidV$ with $\sqrt{\frac{3}{4}\ka\al}$, which should be equal according to the discussion in Section \ref{ss:ComEnvApp}. In the table, the efficiency parameter $\ka$ is estimated in two ways: $\ka_\text{1d}\al$ is estimated from integrating the numerical 1D fluid profiles at $t=1000/\Tc$, while $\ka_\text{Esp}$ comes from the function $\ka(\vw,\alpha)$ given in the Appendix of Ref.\ \cite{Espinosa:2010hh}, using $\vw$ extracted from 1D simulations at $t=1000/\Tc$. As can be seen, $\fluidV$ from the 3D simulations compares reasonably well to its estimate extracted from the 1D numerical profiles around the time of bubble collision, while the theoretical values are somewhat higher. It is remarkable that such a simple model for the mean square velocity, which omits all details of the bubble collisions, does so well.

\subsection{Power spectra}

\begin{figure*}[tb]
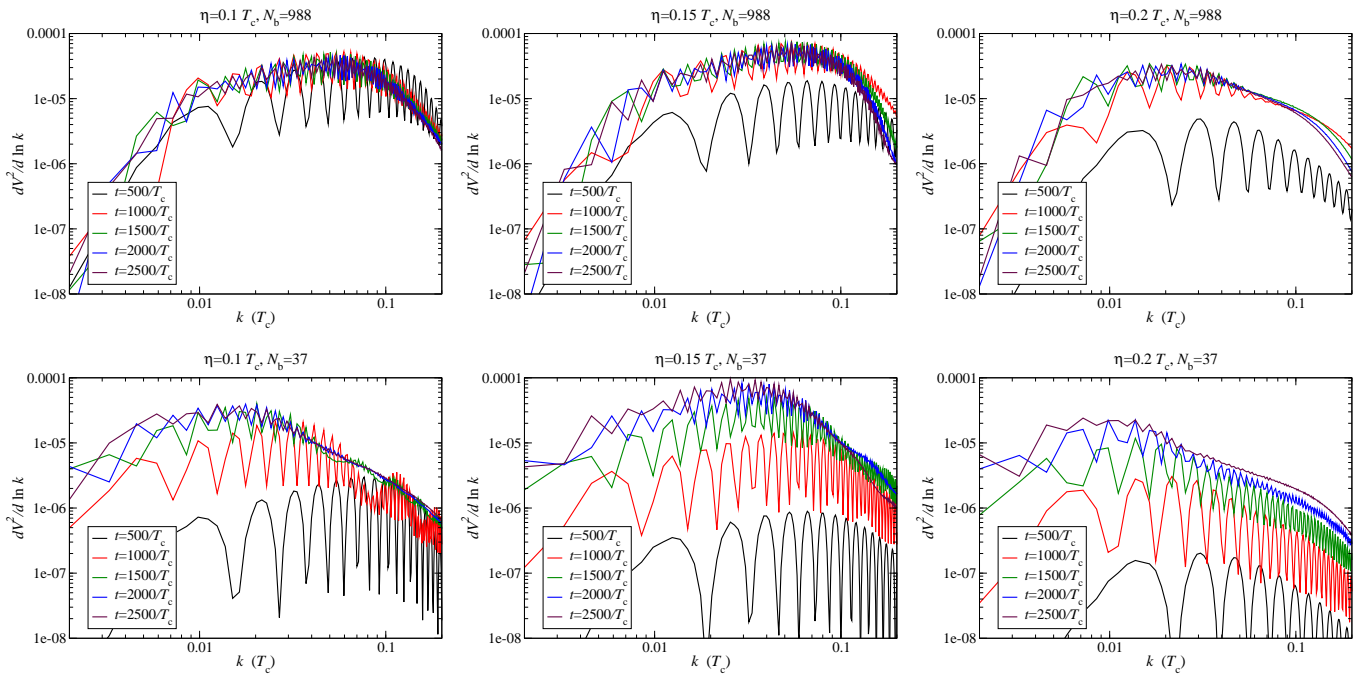

\begin{centering}
\includegraphics[width=0.325\textwidth,clip=true]{velps-0.1-1000.eps}
\hfill
\includegraphics[width=0.325\textwidth,clip=true]{velps-0.15-1000.eps}
\hfill
\includegraphics[width=0.325\textwidth,clip=true]{velps-0.2-1000.eps}\\
\vspace{2ex}
\includegraphics[width=0.325\textwidth,clip=true]{velps-0.1-37.eps}
\hfill
\includegraphics[width=0.325\textwidth,clip=true]{velps-0.15-37.eps}
\hfill
\includegraphics[width=0.325\textwidth,clip=true]{velps-0.2-37.eps}\\

\caption{Velocity power spectra, for weak transitions, at $\eta/T_c = 0.1$, $0.15$ and $0.2$ ($\vw = 0.83$, $0.54$ and $0.44$) for $\Nb = 988$ (top row) and $\Nb=37$ (bottom row). The large oscillations are due to all the bubbles being nucleated at exactly the same time. As in Fig.~\ref{f:GWPSMulti}, we note that the scales are standardised for all the plots, but that the phase transition has not necessarily finished by $2500/Tc$, the time of the latest curve.}

\label{f:VelPSMulti}

\end{centering}
\end{figure*}

\begin{figure*}[tb]
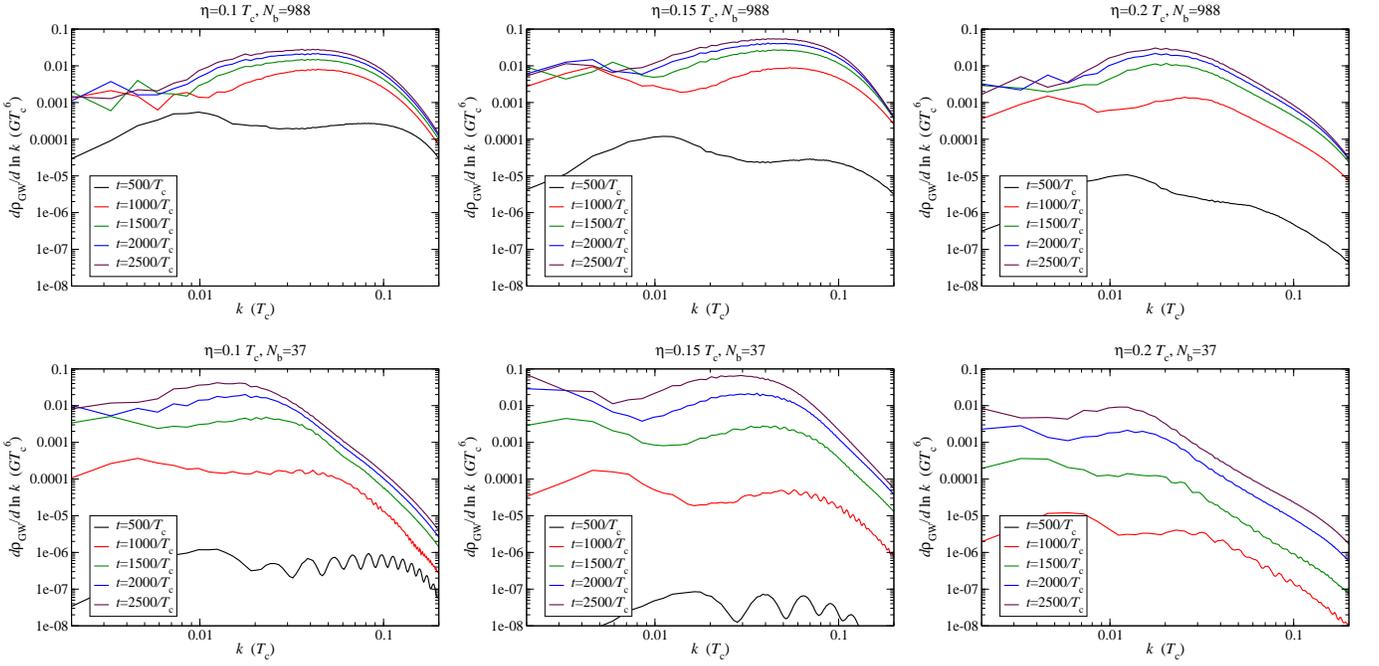

\begin{centering}
\includegraphics[width=0.325\textwidth,clip=true]{gwps-0.1-1000.eps}
\hfill
\includegraphics[width=0.325\textwidth,clip=true]{gwps-0.15-1000.eps}
\hfill
\includegraphics[width=0.325\textwidth,clip=true]{gwps-0.2-1000.eps}\\
\vspace{2ex}
\includegraphics[width=0.325\textwidth,clip=true]{gwps-0.1-37.eps}
\hfill
\includegraphics[width=0.325\textwidth,clip=true]{gwps-0.15-37.eps}
\hfill
\includegraphics[width=0.325\textwidth,clip=true]{gwps-0.2-37.eps}\\

\caption{Gravitational wave power spectra, for weak transitions, at $\eta/T_c = 0.1$, $0.15$ and $0.2$ ($\vw = 0.83$, $0.54$ and $0.44$) for $\Nb = 988$ (top row) and $\Nb=37$ (bottom row). Note that the axes and time intervals are the same for all plots, which means that in some cases the latest ($2500/\Tc$) curve is from before the completion of the phase transition. }

\label{f:GWPSMulti}

\end{centering}
\end{figure*}

In Figs.\ \ref{f:VelPSMulti} and \ref{f:GWPSMulti} we show velocity and gravitational wave power spectra at various times through the simulations, for weak transitions with $\eta/T_c=0.1$, $\eta/T_c=0.15$, and $\eta/T_c=0.2$, where the bubble wall speed is supersonic ($\vw = 0.83$), just subsonic ($\vw = 0.54$), and subsonic ($\vw = 0.44$). The same potential and fluid-field parameters are run with $\Nb=988$ and $\Nb=37$ bubbles, to show the effect of allowing a greater time for the fluid velocity around the expanding bubbles to approach their scaling profiles.  The power spectra develop in characteristic ways in the different phases of the transition, and one can see that if the simulation is stopped too early, a misleading impression of the power spectrum will be obtained.

\subsubsection{Collision phase}

Looking first at the velocity power spectra, the most striking feature is their periodic modulation. This is not a physical feature, and is due to the bubbles being nucleated all at the same time. We have checked that spreading the nucleation times reduces this modulation, and it is not expected to be a feature of the velocity power spectrum of a realistic bubble nucleation distribution in the infinite volume limit. In Appendix \ref{a:BubNucTim} we show the effect of allowing nucleation over a time of about $200/\Tc$.

Once the fluid shells of the nearest pair of bubbles begin to overlap, gravitational waves are generated, at a scale controlled by the size of the bubbles. The overlap of the fluid shells is quickly followed by the collisions of the bubble walls, and gravitational radiation is generated by the scalar field as well. The bubbles continue to grow and to collide, and as a result the length scales of the velocity field and the gravitational radiation get larger (see Fig.~\ref{fig:integralscale}). This effect can be seen in the power spectra, where the curves show a peak moving up and to the left with time.

\begin{figure}
\begin{centering}
\includegraphics[width=0.4\textwidth,clip=true]{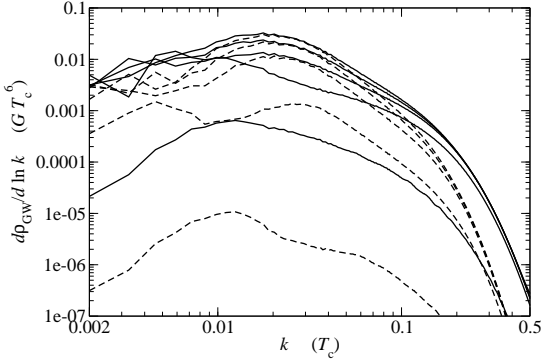}
\caption{Power spectra for $\eta/T_c=0.2$, comparing fluid-only (dashed) and total (solid) GW power at intervals of $500/\Tc$. The power laws visible in the `total GW power' case are dominated by the gradient energy of the scalar field. This source, however, is short-lived. We conjecture that it can be calculated by means of the envelope approximation.}
\label{f:FluVsAll}
\end{centering}
\end{figure}

In our simulations there is generally more energy in the scalar field than in the fluid to begin with, and so the gravitational radiation from the scalar field dominates the early phases (see Fig.~\ref{f:FluVsAll}).  However, when scaled to a real deflagration or detonation in the early universe, most of the latent heat of the transition goes to the fluid, and the radiation from the scalar field can be neglected. It is only in the case of a runaway bubble wall that the scalar field takes most of the latent heat.  We discuss the scaling to real transitions in Section \ref{ss:Extrapolate}, and we plan to study runaway transitions elsewhere.

However, it is interesting to study the difference between a fluid-only gravitational wave power spectrum, and one sourced by both fluid and field (Fig.~\ref{f:FluVsAll}).
There one sees evidence of a $k^{-1}$ power spectrum in arising from the scalar field during the collision phase (solid lines), which is later dominated by the gravitational waves from the fluid. As the scalar field energy density is confined to a thin shell, it is reasonable to suppose that its contribution can be adequately computed in the envelope approximation in the collision phase. We will investigate this conjecture elsewhere.

\subsubsection{Acoustic phase}

Eventually, the low-temperature phase spreads throughout the volume, the scalar field domain walls disappear, and fluid velocity perturbations are left behind.  We call this the acoustic phase of the transition, as the fluid perturbations are primarily compressive (longitudinal) modes (see Appendix \ref{a:TraVelNeg}). During the acoustic phase, the length scale of the fluid perturbations and the gravitational waves remains constant.

For the simulations with fewer bubbles ($\Nb=37$), we see from the lower row of Fig.~\ref{f:VelPSMulti} that the envelope of the velocity power spectrum has an approximate power-law envelope beyond the peak. This power-law envelope is also visible at $\Nb = 988$ at $\eta/\Tc=0.2$, where the bubble wall speed is lower. In both cases at $\eta/\Tc=0.2$, the bubbles expand longer before collision, and we expect the velocity field to be closer to the asymptotic form. We note that the power law is approximately $k^{-1}$ at $\eta/\Tc=0.2$, and appears steeper for lower couplings.  However, we are not confident that we have reached the asymptotic form for bubble wall speeds above $\vw = 0.44$.

At low wavenumbers, the velocity power spectrum behaves as a power of $k$, and arguments based on the analyticity properties of the Fourier transform of a longitudinal vector field in Ref.~\cite{Caprini:2007xq} show that it should go as $k^5$. This is just visible in the first few bins of the simulations with $\Nb=988$. Larger simulations are required to properly check the long-wavelength behaviour.

We see that the gravitational wave power spectrum grows linearly with time in the acoustic phase, maintaining its shape, except at the lowest wavenumbers. A power-law behaviour can be seen emerging beyond the peak, especially in the simulations with $\Nb = 37$. 
The power-law is approximately $k^{-3}$ for the weak deflagrations at $\eta/\Tc=0.2$ and $\eta/\Tc=0.4$ (the power spectra for the latter can be found in the supplementary material for this work) for both $\Nb = 988$ and $\Nb = 37$, which gives us confidence that we are close to the true power law. 
However, a power-law can be seen only for $\Nb = 37$ for $\eta \le 0.15$.  Without further simulations at larger $\Rbc$ we cannot properly determine the long-wavelength behaviour of the growing acoustic phase power spectrum in these cases.

\begin{figure}[t]
\begin{centering}
\includegraphics[width=0.4\textwidth,clip=true]{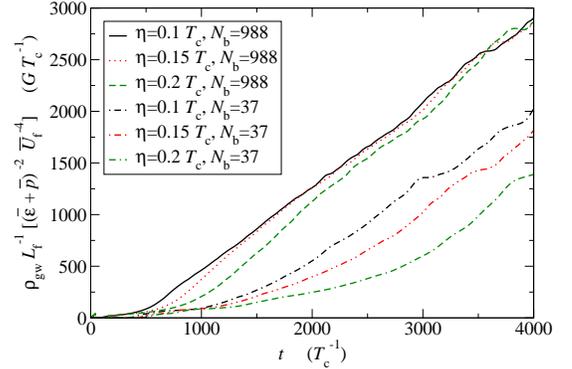}\\
\caption{\label{fig:timevolGW} Time series of $\rho_\text{GW} L_{\mathrm{f}}^{-1} [(\bar \ep+ \bar p)^{-2}\fluidV^{-4}]$, showing the evolution of the gravitational wave energy density relative to an estimate of the square of the final fluid shear stresses. We take the fluid length scale $L_\mathrm{f}$ to be the integral scale $\xi_\mathrm{f}$. Some oscillation about the constant curve is caused by long-wavelength sloshing of the fluid or the infrared behaviour of the gravitational wave power, discussed later, but the striking feature is the scalable linearity of the signal across a factor of three for $R_*$. Only fluid contributions to the gravitational wave power are included here. The early-times steep growth is best explained by the violent behaviour when the two shocks overlap. This phase is not well explained by our random velocity field model.
}
\end{centering}
\end{figure}

We argued earlier that the gravitational wave density parameter $\Omega_\text{GW} = \rho_\text{GW}/\bar\epsilon$ is proportional to $\Rfluid$, the fluid velocity length scale,  and the square of the volume-averaged fluid energy density $(\bar\epsilon + \bar p)^2 {\overline U}_\mathrm{f}^4$. We plot the scaled gravitational wave energy density in Figure~\ref{fig:timevolGW}. This plot shows nicely parallel, linear growth of gravitational wave power when rescaled by these quantities at late times.  The coincidence of the slopes is greatly improved over the equivalent figure in Ref.~\cite{Hindmarsh:2013xza}, thanks to the larger simulation volumes, longer run times, and above all the replacement of the average bubble separation at collision by the fluid integral scale.

The improved coincidence of the slopes is one of the major results of the paper.  It establishes the existence of an $\mathrm{O}(1)$ parameter $8\pi\OmGWscaled$ for a wide range of relevant transitions, and shows that the gravitational wave energy density from a phase transition can be understood in terms of simple features of the velocity field created by the dynamics of the bubble collision.

\subsection{Extrapolating to a real phase transition}
\label{ss:Extrapolate}

Our simulations are necessarily limited in volume, duration, and resolution. We now discuss how they can be extrapolated to the real universe.  In particular we would like to extrapolate the gravitational radiation power spectrum, expressed as a fraction of the critical density. 

There are three physical length scales in the system: the average bubble separation $\Rbc$, the size of the initial bubble of the broken phase $\Rc$, 
and the bubble wall width $\CorLen$.  They are all set by the dimensional scale of the effective potential, which one can chose to be the critical temperature $\Tc$, and various combinations of the dimensionless couplings $\ga$, $\cubPar$ and $\la$.  In a real transition, the average bubble separation is much larger than the wall width because of the exponential factor in the tunnelling rate, whose argument is set by the ratio of the energy of the critical bubble (see Ref.~\cite{Kirzhnits:1976ts}) to the critical temperature. This is generally a large number.

There are also two physical time scales to consider: the lifetime of the fluid flow $\tLife$, and the duration of the phase transition, which is of order $\be^{-1}$, the inverse of the tunnelling rate parameter~(\ref{e:TunRat}).  The duration of the phase transition is also of order $\Rbc/\vw$, the time it takes for bubbles of average separation to collide.

Finally, there are also scales set by the background cosmology: the Hubble rate at the phase transition $\Hc$ (and the Hubble length), and the gravitational constant $G$. The Hubble rate, the gravitational constant and the critical temperature $\Tc$ are related via the Friedmann equation. Our simulations are performed in a Minkowski background, as the duration of the transition is assumed to be comparable to the Hubble time. Therefore $\Tc$ and $G$ can be chosen independently. The role of $G$ is purely to set the scale of of the gravitational perturbations. As mentioned earlier, we use units $\Tc = 1$ and $G=1$.

The observable of interest is the gravitational wave power spectrum, expressed as a fraction of the total density~(\ref{e:GWPowSpe}). It is clear from that formula that the relevant scales are the fluid flow length scale (set by the average bubble separation), the fluid flow lifetime, and the Hubble rate. The power spectrum is determined by the ratios of the fluid flow lifetime to the Hubble time, and the fluid flow scale to the Hubble scale. The role of the bubble wall width is to provide a short-distance cut-off on the power spectrum. A physical transition has $\CorLen < \Rc \ll \Rbc$, with $\Hc\Rbc$ of order 
$10^{-2}$ at the electroweak scale.
Our simulations assume that the bubble separation is much less than the Hubble length, and that the transition rate is much larger than the Hubble rate, so that expansion can be neglected.

The fluid flow lifetime affects only the amplitude of the acoustically generated gravitational waves. In our simulations one sees that after the transition has completed, the power spectrum grows linearly with time while maintaining its shape.
Hence, apart from a trivial scaling, the relevant parameter for the gravitational wave power spectrum is the fluid flow scale, provided that the wall width and the critical bubble size are much less that the bubble separation. 
The effect of too large a ratio $\CorLen/\Rbc$ is that there is insufficient dynamic range to observe the power law behaviour of the power spectrum; the effect of too large a ratio $\Rc/\Rbc$ is that there is insufficient time for the fluid flow to approach its asymptotic self-similar profile, which results in too low a value for $\fluidV$. It also tends to obscure the power law behaviour. 
We have seen in our simulations that the ratio $\CorLen/\Rbc$ needs to be of order $10^{-3}$ in order to reliably distinguish the power law.  
Given our computing resources, this means we are not able to determine the shape of the power spectrum at wave numbers much less than the peak.

\begin{figure}
\begin{centering}
\includegraphics[width=0.4\textwidth,clip=true]{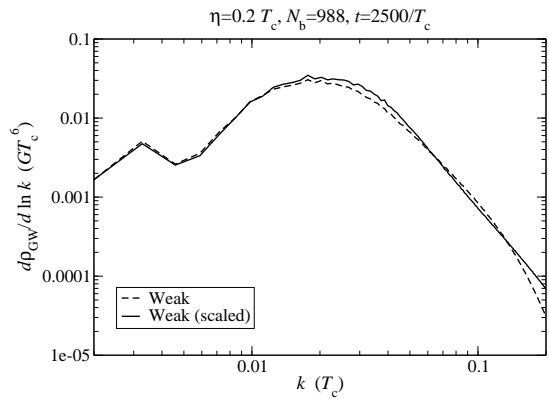}
\caption{Power spectra with $\Nb=988$ comparing the weak phase transition parameters and friction $\eta/T_c=0.2$ with the results from an equivalent run with the scaled parameters. For clarity, only the power spectra at the end of the phase transition ($t=2500/\Tc$) are shown. Note the $y$-axis scale is different to that used in Fig.~\ref{f:GWPSMulti}, in order to highlight the differences between the two power spectra.}
\label{fig:scaledcomp}
\end{centering}
\end{figure}

In order to test the approach to physical ratios we should explore a scaling of the parameters which shrinks the ratios $\CorLen/\Rbc$ and $\Rc/\Rbc$
to zero. Such a scaling was given in Eq.~(\ref{scaling}). Its only effect is to alter the width and surface tension of the bubble wall, and hence shrink the size of the critical bubble and the bubble wall width independently of the bubble separation. 
We carried out a simulation scaled with $r=2$ (so that the bubble wall was half the width) and parameters given in Table \ref{t:SimParsRuns} corresponding to the deflagration, 
and compare the resulting gravitational wave power spectra in Figure~\ref{fig:scaledcomp}.
The power spectra are substantially similar, but the $k^{-3}$ power law is clearer in the scaled run where $\CorLen/\Rbc$ is smaller.  This is consistent with our discussion above, and lends further confidence to our identification of the index of the power law in this case.

Note that the scaling of Eq.~(\ref{scaling}) also reduces the surface tension $\si$ as it reduces the bubble wall width, and hence the relative contribution of the scalar field to the total gravitational wave source tensor (\ref{e:TauDef}), as the following argument makes clear. 
 The scalar field's source tensor $\Tfield$  is proportional to the product of $\si$ with the area per unit volume of the phase boundary, and the area per unit volume is at most of order $1/\Rbc$, which is unaffected by the scaling.  Hence 
 $\Tfield \to r^{-1}\Tfield$. 
 At the same time, the scale of the fluid source tensor $\Tfluid$ is set by the latent heat of the transition ${\cal L}$, which is independent of $r$.  
Hence the the relative importance of the scalar field to the fluid goes as $\CorLen/\Rbc$ as it decreases towards physical values.

This is consistent with the argument given in the introduction that the scalar field contributes negligibly to the gravitational waves, as the ratio of the energy in the scalar field to the energy in the fluid goes as the ratio of the volume in the phase boundary to the total volume, which is  $\CorLen/\Rbc$.

This argument assumes that the bubble walls travel at constant speed, so that the effective surface energy is constant.  Hence if the bubble walls are weakly coupled to the plasma, they can continue to accelerate until they collide~\cite{Bodeker:2009qy}. In this ``run-away'' scenario, scalar fields can contribute importantly to the gravitational radiation.

\section{Discussion and conclusions}

In this paper we have reported on new numerical simulations of the production of gravitational radiation at a first order phase transition in the early universe.
Following standard methods, we model the contents of the universe as a scalar order parameter coupled to a relativistic fluid, with a thermal effective potential~(\ref{e:ScaPot}) and dissipative coupling~(\ref{e:CouTer}). This model captures the essential physics of the transition, which proceeds by the nucleation and growth of bubbles of the low temperature phase. 

The most important parameters of the transition are the latent heat density relative to the total energy density $\strengthPar{T}$, which characterises the strength of the transition, the bubble wall velocity $\vw$, which is determined by $\strengthPar{T}$ and the field-fluid coupling $\eta$, and the bubble nucleation rate parameter $\be$, which determines the average bubble separation.

Most of our simulations are carried out at $\strengthPar{T} \simeq 0.01$, the order of magnitude expected at an electroweak transition, from which we can extrapolate to other values.  We check our extrapolation with a smaller number of simulations at $\strengthPar{T} \simeq 0.1$, and with a scaling argument which changes parameters in the potential without affecting $\strengthPar{T}$.  We simulate	 for a range of phase boundary speeds $\vw$, covering deflagrations and detonations.  Instead of fixing the bubble nucleation rate parameter $\be$, we directly fix the average bubble separation $\Rbc$ by nucleating $\Nb = V/\Rbc^3$ bubbles simultaneously.

We concentrate on the gravitational waves generated by the fluid motion, as the vast majority of the latent heat of the transition is transformed into thermal and kinetic energy of the fluid. 
We show that the gravitational wave density parameter~(\ref{e:OmgwEqn}) is proportional to the fourth power of the mean square fluid velocity, the ratio of lifetime of the source to the Hubble time, and the ratio of length scale of the source to the Hubble length. 

Our results confirm those of a more limited set of simulations reported in Ref.~\cite{Hindmarsh:2013xza}. The fluid kinetic energy is mostly in the form of sound waves generated by the compression or rarefaction of the fluid around the advancing phase boundary. Some rotational flow is generated by the collisions, but at a subdominant level. 
The sound waves remain for as long as we simulate, long after the phase transition completes.  

It was shown in Ref.~\cite{Hindmarsh:2013xza} that when viscosity is included 
the viscous damping time is much longer than the Hubble time  for most phase transitions of interest.
It was argued that the lifetime of the source, the shear stress generated by the sound waves, is approximately the Hubble time. In this paper we detail the calculation which shows that the lifetime parameter $\tLife$, controlled by the decay and decorrelation of the shear stresses,  is in fact exactly the Hubble time.

The length scale of the source, approximately the average bubble separation in Ref.~\cite{Hindmarsh:2013xza}, is here measured directly from the fluid flow.
With this refinement, we show that the proportionality constant $\OmGWscaled$ in the gravitational wave density parameter equation~(\ref{e:OmgwEqn}) varies little between phase transitions with different strength and bubble wall speeds. Indeed, our measurements show that $8\pi\OmGWscaled = 0.8 \pm 0.1$, where the uncertainly is the root mean square fluctuation between simulations.

Our new simulations are carried out on larger lattices, and give a wider dynamic range between the physical scales set by the average bubble separation and the bubble wall width. We further widen the dynamic range by nucleating all bubbles at the same time, at the slight cost of introducing ``ringing'' in the velocity power spectrum.  With the increased dynamic range we are able to establish clear power laws for both velocity and gravitational wave power spectra between the physical scales. 
For the transitions with $\vw = 0.44$ or below, they are $k^{-1}$ and $k^{-3}$ respectively, where $k$ is the wave number, and steeper for the transitions with higher bubble wall speeds. In order to discern these power laws, we show it is important that the fluid velocity profile around the advancing bubble wall has sufficient time to approach its asymptotic self-similar form.

The $k^{-3}$ (or steeper) power law for gravitational waves contrasts with the prediction of $k^{-1}$ from the standard envelope approximation, which assumes that all the energy in the system is concentrated in a thin shell at the bubble wall, and that the radiation is produced only when the shells interact. We see signs of a $k^{-1}$ power spectrum generated by the scalar field in the initial phase of bubble collision, but this component is subdominant in our simulations, and would be completely negligible when extrapolated to the scale separation in a thermal phase transition.

The envelope approximation generically predicts far less gravitational radiation than is actually produced. This under-prediction stems from the incorrect modelling of the source as being the colliding bubble walls.  Instead, the main source is the overlapping sound waves which are left behind after the transition has completed.  We argued in \cite{Hindmarsh:2013xza} that this means that the gravitational wave energy density is boosted by the ratio of the lifetime parameter of the shear stress to the duration of the collision, which goes parametrically as $(\vw\Rfluid\Hc)^{-1}$.  In this paper we studied the numerical factor in this ratio by a careful comparison of the quantities in the envelope approximation formula (\ref{e:EnvAppFor}) to the acoustic generation formula (\ref{e:OmgwEqn}).  We show that the numerical factor is of order unity, and hence we can confirm that the gravitational wave signal is boosted by the ratio of the Hubble time to the phase transition duration, which is two orders of magnitude or more for a typical first order electroweak transition \cite{Hogan:1984hx,Enqvist:1991xw}.

Our simulations shed new light on gravitational waves from phase transitions in the early universe.  They show that the envelope approximation needs to be replaced, both as a model and as a formula.  Instead, we should model the gravitational wave generation in terms of overlapping sound waves.

With this new ``acoustic'' model of gravitational wave generation, we have developed a quantitative understanding of the gravitational wave density parameter (\ref{e:OmgwEqn}), as a function of the mean square fluid velocity and the mean bubble separation.  We can estimate the mean fluid velocity from hydrodynamic considerations \cite{Espinosa:2010hh}, and the mean bubble separation from the nucleation rate parameter $\be$ \cite{Enqvist:1991xw}.
We have numerically determined that the gravitational wave power spectrum is a power law on the high wavenumber side of the peak, and shown that it is steeper than the $k^{-1}$  indicative of a vacuum transition.  Hence potential future observations of such a gravitational wave spectrum will allow us to distinguish between a thermal and a vacuum transition. 

Much remains to be done. We noted that we need larger simulations to trace out the shape of the power spectrum at wavenumbers lower than the peak value, and to determine the index of the power spectrum for the transitions with faster bubble walls. They may also help in the search for bubble wall instabilities identified in~\cite{Link:1992dm,Huet:1992ex,Megevand:2013yua}.
We also need to develop a theoretical understanding of the shape or the power spectrum, and most importantly making accurate quantitative predictions for future gravitational wave observatories. 

\appendix

\section{Gravitational wave production in an expanding universe}
\label{s:GWExpUni}
In this appendix we show how our discussion of the generation of gravitational waves in Section \ref{s:TheGWGen} is modified in an expanding radiation-dominated cosmology. We write the metric 
\ben
g_{\mu\nu} = a^2(\eta)(\eta_{\mu\nu} + h_{\mu\nu}),
\label{e:FLRWmet}
\een
with $\eta$ representing conformal time. In the acoustic phase, after the scalar field has reached its ground state, the energy-momentum tensor takes the standard ideal fluid form
\ben
{T^\mu}_\nu = (\ep + p)U^\mu U_\nu + p \de^\mu_{\;\nu},
\een
with $U^2 = -1$. If we write
\ben
U^0 = W/a, \quad U^i = W V^i/a,
\een
then $W^2 = 1/(1 - V^2)$, and $V^i$ resembles the Minkowski space 3-velocity.

Indeed, it can be shown \cite{Brandenburg:1996fc} that the relativistic fluid equations in the radiation era are exactly the same as the Minkowski space equations, when the fluid variables are appropriately scaled. So writing
\ben
\tilde{E} = \frac{a^4}{a_*^4} E, \quad \tilde{Z}_i = \frac{a^4}{a_*^4} Z_i  , 
\een
with $a_*$ some reference scale factor,
the equations for the tilde fields and $V^i$ are identical to the fluid parts of Eqs.~(\ref{eqE}) and (\ref{eqZ}), with $t$ interpreted as conformal time, and $x^i$ as comoving coordinates.  Hence our simulations need no adaptation for an expanding universe in the acoustic phase, where only fluid variables are active, and the expanding universe energy-momentum tensor can be obtained by multiplying by appropriate powers of the scale factor $a$. For example, 
the relevant quantity for gravitational wave generation is the fluid source tensor with both indices down (\ref{e:TauDef}). This can be written 
\ben
\Tfluid_{ij}(\bk,\eta) = 
\frac{a_*^4}{a^2(\eta)}\tilde\tau^\text{f}_{ij}(\bk,\eta),
\een
where we choose $a_*$ to be the scale factor at the phase transition time $\eta_*$, and $\tilde\tau^\text{f}_{ij}$ represents the source tensor obtained from the fluid evolution in scaled coordinates $V^i$, $\tilde{E}$ and $\tilde{Z}_i$.

\begin{widetext}

We then see that the FLRW version of Eq.~(\ref{e:UETCmod}) is 
\ben
\label{e:FRWUETCmod}
\Pi^2(k,t_1,t_2) \simeq \frac{a_*^8}{a^2(\eta_1)a^2(\eta_2)}[(\bar\ep+\bar p)\fluidV^2]^2\Rfluid^3\tilde\Pi^2(k\Rfluid,k\et_1,k\et_2),
\een
where $\bar\ep$, $\bar p$, $\fluidV^2$ and $\tilde\Pi$ are the values of the energy density pressure and mean square velocity in  the scaled fields, $\Rfluid$ is the comoving length scale of the fluid perturbations, and $\tilde\Pi^2$ is the same dimensionless function as in (\ref{e:UETCmod}). We emphasise that $\bar\ep$, $\bar p$, $\fluidV^2$ and $\tilde\Pi$ are those measured in our Minkowski space numerical simulations.

In the metric (\ref{e:FLRWmet}), the solution to the radiation era field equation for the tensor mode (\ref{e:GWsol}) is modified to 
\ben
h_{ij} (\mathbf{k},\eta) = (16\pi G)\la_{ij,kl}(\bk) \int_0^\eta d\eta' \frac{\sin[k(\eta-\eta')]}{k} \frac{a(\eta')}{a(\eta)} \ta_{kl}(\bk,\eta'),
\een
and the definition of the gravitational wave energy density power spectrum (\ref{e:GWPowSpeDef}) becomes   
\ben
\frac{d\rho_{\rm gw}}{d\log k} = \frac{1}{32\pi G}\frac{k^3}{2\pi^2}\frac{\SpecDen{\dot h}}{a^2}.
\een
Writing $x = k(\eta_1+\eta_2)/2$ and $z = k(\eta_1-\eta_2)$, and using the radiation era scale factor $a(\eta) = (\eta/\eta_*)a_*$, 
the spectral density of $\dot{h}$ can be written 
\ben
\SpecDen{\dot h}(k,\eta) = \frac{a_*^6}{a^4(\eta)} \left[16\pi G(\bar\ep+\bar p)\fluidV^2\right]^2   \Rfluid^3 \int_{k\eta_*}^{k\eta} dx \int_{z_-}^{z_+} dz \frac{\eta_*^2}{x^2 - z^2/4} \frac{\cos(z)}{2} \tilde\Pi^2(k\Rfluid,z,x),
\label{e:FRWSpeDenExpA}
\een
where $z_\pm = \pm 2(x - k\eta_*)$ for $x \le k(\eta+\eta_*)/2$ and $z_\pm = \pm 2(k\eta -x )$ for $x > k(\eta+\eta_*)/2$.
.
Making the assumptions that the autocorrelation time of the fluid perturbations is small compared with the Hubble time, so that the integrand is negligible when $z$ approaches $x$, and that the correlations are approximately stationary (independent of $x$) over the domain of integration we find
\ben
\SpecDen{\dot h}(k,\eta) \simeq \frac{a_*^4}{a^4(\eta)} \left[16\pi G(\bar\ep+\bar p)\fluidV^2\right]^2  (a_* \eta_*)(a_* k^{-1}) \Rfluid^3   \int_{-\infty}^\infty dz  \frac{\cos(z)}{2} \tilde\Pi^2(k\Rfluid,z).
\label{e:FRWSpeDenExpB}
\een
We see that the expression has the same form as Eq.~(\ref{e:SpeDenExpA}), 
but with the Minkowski time replaced by $a_* \eta_*$ (the physical Hubble time) and the Minkowski wavenumber replaced by its  physical value at time $\eta_*$. We also see the correct redshift factor for the gravitational radiation.

Hence we can immediately write down the analogue of Eq.~(42)
\bea
\frac{d \OmGW(k)}{d \ln(k)} &=&  3(1+w)^2 \fluidV^4  (\Hc\Rfluid^*) \frac{(k\Rfluid)^3}{2\pi^2} \SpecDenGW(k\Rfluid),
\eea
where $\Rfluid^*$ means the physical length scale at $\eta_*$, while $\SpecDenGW(k\Rfluid)$ 
remains its Minkowski space version (\ref{e:NoDimSpecDen}), with $k$ and $\Rfluid$ interpreted as comoving quantities.
We therefore learn that the effective lifetime of the source for gravitational waves is precisely the Hubble time.

Note that this effective lifetime is not the lifetime of the acoustic waves themselves, whose density perturbation continues to oscillate with constant amplitude in the absence of dissipation. 
Instead, it appears as a result of a combination of the expansion damping and decorrelation of the shear stress.
To see the effect of the decorrelation, let us consider shear stress correlations behaving as $\tilde\Pi_0^2\cos(z)$, in which case the $z$ integrand in (\ref{e:FRWSpeDenExpB})  would always be positive, representing the largest possible growth rate for the gravitational wave power spectrum.  In this extreme case, representing the effect of expansion damping alone, the factor $(a_* \eta_*)(a_* k^{-1})$ in (\ref{e:FRWSpeDenExpB}) would be replaced by as $(a_* \eta_*)^2 \ln^2(\eta/\eta_*)$.  The decorrelation of the shear stresses cuts off the $z$ integral, removing the logarithms and replacing one factor of the Hubble time with a factor of the wavelength.

\end{widetext}

\section{Simulations with a range of bubble nucleation times}
\label{a:BubNucTim}

In Section~\ref{s:NumRes} we claim that the large oscillations in the
power spectra presented in the main body of the paper are due to the
bubbles being nucleated at the same time. There are many thin fluid shells of the same radial size that contribute to the power spectrum, and the Fourier transforms of both the fluid velocity $V^i$ and unprojected metric perturbations $h_{ij}$ will be in phase for all such shells.
Here we demonstrate that such oscillations are damped when the bubbles are nucleated over a more widely spread period.

We ran a single simulation with our ``weak'' potential parameters and $\eta/T_c=0.2$, nucleating bubbles with a rate parameter $\beta=0.01$. However, we capped the number of bubbles to approximately 1000 (in the end we nucleated 1002), leading to an abrupt cutoff in the exponential growth of the number of bubbles rather than the full double exponential seen in Refs.~\cite{Hindmarsh:2013xza,Enqvist:1991xw}. The last bubble is nucleated shortly before $t=200/\Tc$.

\begin{figure}
\begin{centering}

\includegraphics[width=0.4\textwidth,clip=true]{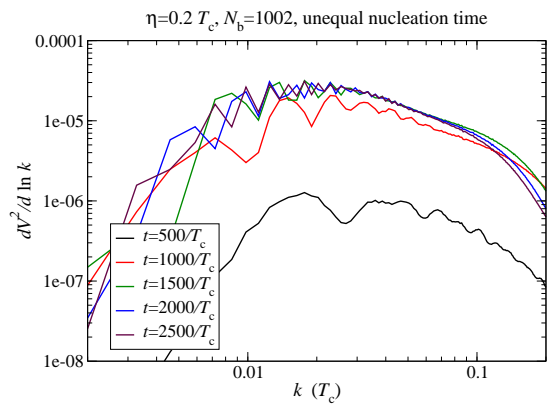}

\caption{\label{f:SpreadNucTime}  Velocity power spectrum with a range of nucleation times. Compared to the similar simulation results in Fig.~\ref{f:VelPSMulti}, top right, the oscillations in the fluid power are significantly more damped. This demonstrates that the strong oscillatory behaviour in the fluid velocity power spectra is due to the equal nucleation time.}
\end{centering}
\end{figure}

While the number of bubbles and the spread of nucleation times will change the parameters we studied in the main body of the paper, they are sufficiently close for the purposes of this appendix. In Fig.~\ref{f:SpreadNucTime} we show the fluid velocity power spectrum (in the same manner as Fig.~\ref{f:VelPSMulti}) and demonstrate that the oscillations are considerably damped compared to the equivalent plot in Fig.~\ref{f:VelPSMulti}, top right.

\section{Transverse modes of the velocity field are negligible}
\label{a:TraVelNeg}

Here, we show that the power in the transverse modes of the fluid velocity is significantly smaller than that in the longitudinal modes, supporting our claim that the fluid perturbations are best characterised as an essentially linear superposition of sound waves.

One way of quantifying this is to study the RMS fluid velocity when only transverse motion is taken into account. We quantify this with $\fluidVmaxperp$ in Table~\ref{t:SimParsRuns}. It can be seen that the transverse modes contribute at most $5-10\%$ to the RMS fluid velocity $\fluidV$. This ratio is greatest for the simulations with the largest $\Nb$ (and hence smallest $\Rbc$), so we would expect in a realistic scenario (with $\Rbc$ considerably larger) that the transverse modes would be very small indeed.

\begin{figure}
\begin{centering}
\includegraphics[width=0.4\textwidth,clip=true]{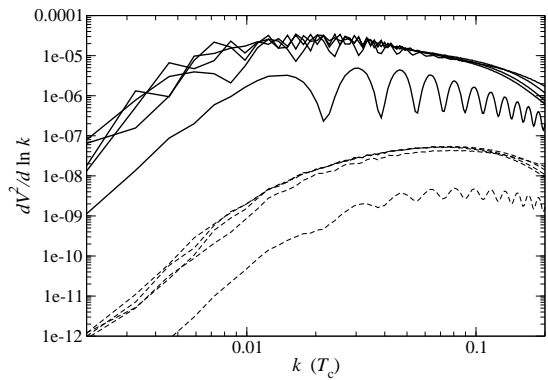}

\caption{\label{fig:divrot} Comparison of longitudinal (solid) and transverse (dashed) fluid velocity power spectra for $\Nb=988$, $\eta/T_c=0.2$. The time intervals are $500/T_\mathrm{c}$ as in Fig.~\ref{f:VelPSMulti} (c.f. top right figure). The transverse part of the fluid velocity power is many orders of magnitude smaller than the longitudinal part and does not play a significant role in gravitational wave production. Note the extended scale compared to Fig.~\ref{f:VelPSMulti}.}
\end{centering}
\end{figure}

The maximum value of $\fluidVperp$ (which is the value quoted in Table~\ref{t:SimParsRuns}) occurs at approximately the same time as the maximum value of $\fluidV$, at around the conclusion of the phase transition before slowly decreasing due to discretisation effects. This behaviour can be seen in Figure~\ref{fig:timevolV}.

Finally, we show in Figure~\ref{fig:divrot} the transverse fluid velocity power spectrum (dashed lines) for a typical simulation alongside the longitudinal power (solid lines). It can be seen that it is significantly smaller than the longitudinal power (a ratio of around $10^{-3}$). 

It has been predicted that turbulent fluid motion would develop after a phase transition such as the one under study in this paper. Fluid turbulence is generally studied in incompressible (i.e.\ rotational) flows, with energy transport from a larger forcing  scale to a smaller dissipation scale, leading to the formation of characteristic power laws. 
No clear power law can be seen in the transverse velocity power spectrum, which leads us to believe that turbulence is not a feature of our simulations.

It has recently been suggested that turbulence can develop in the acoustic perturbations \cite{Kalaydzhyan:2014wca}, giving rise to an inverse cascade (transfer of power to longer scales).  We see no signs of the length scale of the acoustic oscillations changing once the transition is complete, and the velocity power spectrum does not change its form significantly. 

While it is possible that turbulence develops at larger Reynolds numbers than we have access to\footnote{Our largest Reynolds number, which can be estimated from the ratio $\IntSca/\de x$ given that the fluid viscosity is numerical, is about 1000.}, it is clear that it has no significance for gravitational radiation in the relatively weak transitions we study.

\begin{acknowledgments}
Our simulations made use of facilities at the Finnish Centre for Scientific Computing CSC, and the COSMOS Consortium supercomputer (within the DiRAC Facility jointly funded by STFC and the Large Facilities Capital Fund of BIS). KR acknowledges support from the Academy of Finland project 1134018; MH and SH from the Science and Technology Facilities Council (grant number ST/J000477/1). DJW acknowledges support from the Magnus Ehrnrooth Foundation for travel to the University of Sussex, and is supported by the People Programme (Marie Sk{\l}odowska-Curie actions) of the European Union Seventh Framework Programme (FP7/2007-2013) under grant agreement number PIEF-GA-2013-629425.
\end{acknowledgments}

\bibliography{longbubbles}

\begin{thebibliography}{60}%
\makeatletter
\providecommand \@ifxundefined [1]{%
 \@ifx{#1\undefined}
}%
\providecommand \@ifnum [1]{%
 \ifnum #1\expandafter \@firstoftwo
 \else \expandafter \@secondoftwo
 \fi
}%
\providecommand \@ifx [1]{%
 \ifx #1\expandafter \@firstoftwo
 \else \expandafter \@secondoftwo
 \fi
}%
\providecommand \natexlab [1]{#1}%
\providecommand \enquote  [1]{``#1''}%
\providecommand \bibnamefont  [1]{#1}%
\providecommand \bibfnamefont [1]{#1}%
\providecommand \citenamefont [1]{#1}%
\providecommand \href@noop [0]{\@secondoftwo}%
\providecommand \href [0]{\begingroup \@sanitize@url \@href}%
\providecommand \@href[1]{\@@startlink{#1}\@@href}%
\providecommand \@@href[1]{\endgroup#1\@@endlink}%
\providecommand \@sanitize@url [0]{\catcode `\\12\catcode `\$12\catcode
  `\&12\catcode `\#12\catcode `\^12\catcode `\_12\catcode `\%12\relax}%
\providecommand \@@startlink[1]{}%
\providecommand \@@endlink[0]{}%
\providecommand \url  [0]{\begingroup\@sanitize@url \@url }%
\providecommand \@url [1]{\endgroup\@href {#1}{\urlprefix }}%
\providecommand \urlprefix  [0]{URL }%
\providecommand \Eprint [0]{\href }%
\providecommand \doibase [0]{http://dx.doi.org/}%
\providecommand \selectlanguage [0]{\@gobble}%
\providecommand \bibinfo  [0]{\@secondoftwo}%
\providecommand \bibfield  [0]{\@secondoftwo}%
\providecommand \translation [1]{[#1]}%
\providecommand \BibitemOpen [0]{}%
\providecommand \bibitemStop [0]{}%
\providecommand \bibitemNoStop [0]{.\EOS\space}%
\providecommand \EOS [0]{\spacefactor3000\relax}%
\providecommand \BibitemShut  [1]{\csname bibitem#1\endcsname}%
\let\auto@bib@innerbib\@empty
\bibitem [{\citenamefont {Harry}(2010)}]{Harry:2010zz}%
  \BibitemOpen
  \bibfield  {author} {\bibinfo {author} {\bibfnamefont {G.~M.}\ \bibnamefont
  {Harry}} (\bibinfo {collaboration} {LIGO Scientific}),\ }\href {\doibase
  10.1088/0264-9381/27/8/084006} {\bibfield  {journal} {\bibinfo  {journal}
  {Class.Quant.Grav.}\ }\textbf {\bibinfo {volume} {27}},\ \bibinfo {pages}
  {084006} (\bibinfo {year} {2010})}\BibitemShut {NoStop}%
\bibitem [{\citenamefont {Accadia}\ \emph {et~al.}(2009)\citenamefont {Accadia}
  \emph {et~al.}}]{Accadia:2009zz}%
  \BibitemOpen
  \bibfield  {author} {\bibinfo {author} {\bibfnamefont {T.}~\bibnamefont
  {Accadia}} \emph {et~al.},\ }in\ \href {\doibase 10.1142/9789814374552_0313}
  {\emph {\bibinfo {booktitle} {Proceedings, 12th Marcel Grossmann Meeting on
  General Relativity, Paris, France, July 12-18, 2009. Vol. 1-3}}}\ (\bibinfo
  {year} {2009})\ pp.\ \bibinfo {pages} {1738--1742}\BibitemShut {NoStop}%
\bibitem [{\citenamefont {Somiya}(2012)}]{Somiya:2011np}%
  \BibitemOpen
  \bibfield  {author} {\bibinfo {author} {\bibfnamefont {K.}~\bibnamefont
  {Somiya}} (\bibinfo {collaboration} {KAGRA}),\ }\href {\doibase
  10.1088/0264-9381/29/12/124007} {\bibfield  {journal} {\bibinfo  {journal}
  {Class.Quant.Grav.}\ }\textbf {\bibinfo {volume} {29}},\ \bibinfo {pages}
  {124007} (\bibinfo {year} {2012})},\ \Eprint {http://arxiv.org/abs/1111.7185}
  {arXiv:1111.7185 [gr-qc]} \BibitemShut {NoStop}%
\bibitem [{\citenamefont {Abadie}\ \emph {et~al.}(2010)\citenamefont {Abadie}
  \emph {et~al.}}]{Abadie:2010cf}%
  \BibitemOpen
  \bibfield  {author} {\bibinfo {author} {\bibfnamefont {J.}~\bibnamefont
  {Abadie}} \emph {et~al.} (\bibinfo {collaboration} {LIGO Scientific,
  VIRGO}),\ }\href {\doibase 10.1088/0264-9381/27/17/173001} {\bibfield
  {journal} {\bibinfo  {journal} {Class.Quant.Grav.}\ }\textbf {\bibinfo
  {volume} {27}},\ \bibinfo {pages} {173001} (\bibinfo {year} {2010})},\
  \Eprint {http://arxiv.org/abs/1003.2480} {arXiv:1003.2480 [astro-ph.HE]}
  \BibitemShut {NoStop}%
\bibitem [{\citenamefont {Binetruy}\ \emph {et~al.}(2012)\citenamefont
  {Binetruy}, \citenamefont {Bohe}, \citenamefont {Caprini},\ and\
  \citenamefont {Dufaux}}]{Binetruy:2012ze}%
  \BibitemOpen
  \bibfield  {author} {\bibinfo {author} {\bibfnamefont {P.}~\bibnamefont
  {Binetruy}}, \bibinfo {author} {\bibfnamefont {A.}~\bibnamefont {Bohe}},
  \bibinfo {author} {\bibfnamefont {C.}~\bibnamefont {Caprini}}, \ and\
  \bibinfo {author} {\bibfnamefont {J.-F.}\ \bibnamefont {Dufaux}},\ }\href
  {\doibase 10.1088/1475-7516/2012/06/027} {\bibfield  {journal} {\bibinfo
  {journal} {JCAP}\ }\textbf {\bibinfo {volume} {1206}},\ \bibinfo {pages}
  {027} (\bibinfo {year} {2012})},\ \Eprint {http://arxiv.org/abs/1201.0983}
  {arXiv:1201.0983 [gr-qc]} \BibitemShut {NoStop}%
\bibitem [{\citenamefont {Seoane}\ \emph {et~al.}(2013)\citenamefont {Seoane}
  \emph {et~al.}}]{Seoane:2013qna}%
  \BibitemOpen
  \bibfield  {author} {\bibinfo {author} {\bibfnamefont {P.~A.}\ \bibnamefont
  {Seoane}} \emph {et~al.} (\bibinfo {collaboration} {eLISA}),\ }\href@noop {}
  {\  (\bibinfo {year} {2013})},\ \Eprint {http://arxiv.org/abs/1305.5720}
  {arXiv:1305.5720 [astro-ph.CO]} \BibitemShut {NoStop}%
\bibitem [{\citenamefont {Kajantie}\ \emph
  {et~al.}(1996{\natexlab{a}})\citenamefont {Kajantie}, \citenamefont {Laine},
  \citenamefont {Rummukainen},\ and\ \citenamefont
  {Shaposhnikov}}]{Kajantie:1995kf}%
  \BibitemOpen
  \bibfield  {author} {\bibinfo {author} {\bibfnamefont {K.}~\bibnamefont
  {Kajantie}}, \bibinfo {author} {\bibfnamefont {M.}~\bibnamefont {Laine}},
  \bibinfo {author} {\bibfnamefont {K.}~\bibnamefont {Rummukainen}}, \ and\
  \bibinfo {author} {\bibfnamefont {M.~E.}\ \bibnamefont {Shaposhnikov}},\
  }\href {\doibase 10.1016/0550-3213(96)00052-1} {\bibfield  {journal}
  {\bibinfo  {journal} {Nucl.Phys.}\ }\textbf {\bibinfo {volume} {B466}},\
  \bibinfo {pages} {189} (\bibinfo {year} {1996}{\natexlab{a}})},\ \Eprint
  {http://arxiv.org/abs/hep-lat/9510020} {arXiv:hep-lat/9510020 [hep-lat]}
  \BibitemShut {NoStop}%
\bibitem [{\citenamefont {Kajantie}\ \emph
  {et~al.}(1996{\natexlab{b}})\citenamefont {Kajantie}, \citenamefont {Laine},
  \citenamefont {Rummukainen},\ and\ \citenamefont
  {Shaposhnikov}}]{Kajantie:1996mn}%
  \BibitemOpen
  \bibfield  {author} {\bibinfo {author} {\bibfnamefont {K.}~\bibnamefont
  {Kajantie}}, \bibinfo {author} {\bibfnamefont {M.}~\bibnamefont {Laine}},
  \bibinfo {author} {\bibfnamefont {K.}~\bibnamefont {Rummukainen}}, \ and\
  \bibinfo {author} {\bibfnamefont {M.~E.}\ \bibnamefont {Shaposhnikov}},\
  }\href {\doibase 10.1103/PhysRevLett.77.2887} {\bibfield  {journal} {\bibinfo
   {journal} {Phys.Rev.Lett.}\ }\textbf {\bibinfo {volume} {77}},\ \bibinfo
  {pages} {2887} (\bibinfo {year} {1996}{\natexlab{b}})},\ \Eprint
  {http://arxiv.org/abs/hep-ph/9605288} {arXiv:hep-ph/9605288 [hep-ph]}
  \BibitemShut {NoStop}%
\bibitem [{\citenamefont {Gurtler}\ \emph {et~al.}(1997)\citenamefont
  {Gurtler}, \citenamefont {Ilgenfritz},\ and\ \citenamefont
  {Schiller}}]{Gurtler:1997hr}%
  \BibitemOpen
  \bibfield  {author} {\bibinfo {author} {\bibfnamefont {M.}~\bibnamefont
  {Gurtler}}, \bibinfo {author} {\bibfnamefont {E.-M.}\ \bibnamefont
  {Ilgenfritz}}, \ and\ \bibinfo {author} {\bibfnamefont {A.}~\bibnamefont
  {Schiller}},\ }\href {\doibase 10.1103/PhysRevD.56.3888} {\bibfield
  {journal} {\bibinfo  {journal} {Phys.Rev.}\ }\textbf {\bibinfo {volume}
  {D56}},\ \bibinfo {pages} {3888} (\bibinfo {year} {1997})},\ \Eprint
  {http://arxiv.org/abs/hep-lat/9704013} {arXiv:hep-lat/9704013 [hep-lat]}
  \BibitemShut {NoStop}%
\bibitem [{\citenamefont {Csikor}\ \emph {et~al.}(1999)\citenamefont {Csikor},
  \citenamefont {Fodor},\ and\ \citenamefont {Heitger}}]{Csikor:1998eu}%
  \BibitemOpen
  \bibfield  {author} {\bibinfo {author} {\bibfnamefont {F.}~\bibnamefont
  {Csikor}}, \bibinfo {author} {\bibfnamefont {Z.}~\bibnamefont {Fodor}}, \
  and\ \bibinfo {author} {\bibfnamefont {J.}~\bibnamefont {Heitger}},\ }\href
  {\doibase 10.1103/PhysRevLett.82.21} {\bibfield  {journal} {\bibinfo
  {journal} {Phys.Rev.Lett.}\ }\textbf {\bibinfo {volume} {82}},\ \bibinfo
  {pages} {21} (\bibinfo {year} {1999})},\ \Eprint
  {http://arxiv.org/abs/hep-ph/9809291} {arXiv:hep-ph/9809291 [hep-ph]}
  \BibitemShut {NoStop}%
\bibitem [{\citenamefont {D'Onofrio}\ \emph {et~al.}(2014)\citenamefont
  {D'Onofrio}, \citenamefont {Rummukainen},\ and\ \citenamefont
  {Tranberg}}]{DOnofrio:2014kta}%
  \BibitemOpen
  \bibfield  {author} {\bibinfo {author} {\bibfnamefont {M.}~\bibnamefont
  {D'Onofrio}}, \bibinfo {author} {\bibfnamefont {K.}~\bibnamefont
  {Rummukainen}}, \ and\ \bibinfo {author} {\bibfnamefont {A.}~\bibnamefont
  {Tranberg}},\ }\href {\doibase 10.1103/PhysRevLett.113.141602} {\bibfield
  {journal} {\bibinfo  {journal} {Phys.Rev.Lett.}\ }\textbf {\bibinfo {volume}
  {113}},\ \bibinfo {pages} {141602} (\bibinfo {year} {2014})},\ \Eprint
  {http://arxiv.org/abs/1404.3565} {arXiv:1404.3565 [hep-ph]} \BibitemShut
  {NoStop}%
\bibitem [{\citenamefont {Carena}\ \emph {et~al.}(1996)\citenamefont {Carena},
  \citenamefont {Quiros},\ and\ \citenamefont {Wagner}}]{Carena:1996wj}%
  \BibitemOpen
  \bibfield  {author} {\bibinfo {author} {\bibfnamefont {M.~S.}\ \bibnamefont
  {Carena}}, \bibinfo {author} {\bibfnamefont {M.}~\bibnamefont {Quiros}}, \
  and\ \bibinfo {author} {\bibfnamefont {C.}~\bibnamefont {Wagner}},\ }\href
  {\doibase 10.1016/0370-2693(96)00475-3} {\bibfield  {journal} {\bibinfo
  {journal} {Phys.Lett.}\ }\textbf {\bibinfo {volume} {B380}},\ \bibinfo
  {pages} {81} (\bibinfo {year} {1996})},\ \Eprint
  {http://arxiv.org/abs/hep-ph/9603420} {arXiv:hep-ph/9603420 [hep-ph]}
  \BibitemShut {NoStop}%
\bibitem [{\citenamefont {Delepine}\ \emph {et~al.}(1996)\citenamefont
  {Delepine}, \citenamefont {Gerard}, \citenamefont {Gonzalez~Felipe},\ and\
  \citenamefont {Weyers}}]{Delepine:1996vn}%
  \BibitemOpen
  \bibfield  {author} {\bibinfo {author} {\bibfnamefont {D.}~\bibnamefont
  {Delepine}}, \bibinfo {author} {\bibfnamefont {J.}~\bibnamefont {Gerard}},
  \bibinfo {author} {\bibfnamefont {R.}~\bibnamefont {Gonzalez~Felipe}}, \ and\
  \bibinfo {author} {\bibfnamefont {J.}~\bibnamefont {Weyers}},\ }\href
  {\doibase 10.1016/0370-2693(96)00921-5} {\bibfield  {journal} {\bibinfo
  {journal} {Phys.Lett.}\ }\textbf {\bibinfo {volume} {B386}},\ \bibinfo
  {pages} {183} (\bibinfo {year} {1996})},\ \Eprint
  {http://arxiv.org/abs/hep-ph/9604440} {arXiv:hep-ph/9604440 [hep-ph]}
  \BibitemShut {NoStop}%
\bibitem [{\citenamefont {Laine}\ and\ \citenamefont
  {Rummukainen}(1998)}]{Laine:1998qk}%
  \BibitemOpen
  \bibfield  {author} {\bibinfo {author} {\bibfnamefont {M.}~\bibnamefont
  {Laine}}\ and\ \bibinfo {author} {\bibfnamefont {K.}~\bibnamefont
  {Rummukainen}},\ }\href {\doibase 10.1016/S0550-3213(98)00530-6} {\bibfield
  {journal} {\bibinfo  {journal} {Nucl.Phys.}\ }\textbf {\bibinfo {volume}
  {B535}},\ \bibinfo {pages} {423} (\bibinfo {year} {1998})},\ \Eprint
  {http://arxiv.org/abs/hep-lat/9804019} {arXiv:hep-lat/9804019 [hep-lat]}
  \BibitemShut {NoStop}%
\bibitem [{\citenamefont {Grojean}\ \emph {et~al.}(2005)\citenamefont
  {Grojean}, \citenamefont {Servant},\ and\ \citenamefont
  {Wells}}]{Grojean:2004xa}%
  \BibitemOpen
  \bibfield  {author} {\bibinfo {author} {\bibfnamefont {C.}~\bibnamefont
  {Grojean}}, \bibinfo {author} {\bibfnamefont {G.}~\bibnamefont {Servant}}, \
  and\ \bibinfo {author} {\bibfnamefont {J.~D.}\ \bibnamefont {Wells}},\ }\href
  {\doibase 10.1103/PhysRevD.71.036001} {\bibfield  {journal} {\bibinfo
  {journal} {Phys.Rev.}\ }\textbf {\bibinfo {volume} {D71}},\ \bibinfo {pages}
  {036001} (\bibinfo {year} {2005})},\ \Eprint
  {http://arxiv.org/abs/hep-ph/0407019} {arXiv:hep-ph/0407019 [hep-ph]}
  \BibitemShut {NoStop}%
\bibitem [{\citenamefont {Huber}\ and\ \citenamefont
  {Schmidt}(2001)}]{Huber:2000mg}%
  \BibitemOpen
  \bibfield  {author} {\bibinfo {author} {\bibfnamefont {S.}~\bibnamefont
  {Huber}}\ and\ \bibinfo {author} {\bibfnamefont {M.}~\bibnamefont
  {Schmidt}},\ }\href {\doibase 10.1016/S0550-3213(01)00250-4} {\bibfield
  {journal} {\bibinfo  {journal} {Nucl.Phys.}\ }\textbf {\bibinfo {volume}
  {B606}},\ \bibinfo {pages} {183} (\bibinfo {year} {2001})},\ \Eprint
  {http://arxiv.org/abs/hep-ph/0003122} {arXiv:hep-ph/0003122 [hep-ph]}
  \BibitemShut {NoStop}%
\bibitem [{\citenamefont {Huber}\ \emph {et~al.}(2006)\citenamefont {Huber},
  \citenamefont {Konstandin}, \citenamefont {Prokopec},\ and\ \citenamefont
  {Schmidt}}]{Huber:2006wf}%
  \BibitemOpen
  \bibfield  {author} {\bibinfo {author} {\bibfnamefont {S.~J.}\ \bibnamefont
  {Huber}}, \bibinfo {author} {\bibfnamefont {T.}~\bibnamefont {Konstandin}},
  \bibinfo {author} {\bibfnamefont {T.}~\bibnamefont {Prokopec}}, \ and\
  \bibinfo {author} {\bibfnamefont {M.~G.}\ \bibnamefont {Schmidt}},\ }\href
  {\doibase 10.1016/j.nuclphysb.2006.09.003} {\bibfield  {journal} {\bibinfo
  {journal} {Nucl.Phys.}\ }\textbf {\bibinfo {volume} {B757}},\ \bibinfo
  {pages} {172} (\bibinfo {year} {2006})},\ \Eprint
  {http://arxiv.org/abs/hep-ph/0606298} {arXiv:hep-ph/0606298 [hep-ph]}
  \BibitemShut {NoStop}%
\bibitem [{\citenamefont {Dorsch}\ \emph {et~al.}(2013)\citenamefont {Dorsch},
  \citenamefont {Huber},\ and\ \citenamefont {No}}]{Dorsch:2013wja}%
  \BibitemOpen
  \bibfield  {author} {\bibinfo {author} {\bibfnamefont {G.}~\bibnamefont
  {Dorsch}}, \bibinfo {author} {\bibfnamefont {S.}~\bibnamefont {Huber}}, \
  and\ \bibinfo {author} {\bibfnamefont {J.}~\bibnamefont {No}},\ }\href
  {\doibase 10.1007/JHEP10(2013)029} {\bibfield  {journal} {\bibinfo  {journal}
  {JHEP}\ }\textbf {\bibinfo {volume} {1310}},\ \bibinfo {pages} {029}
  (\bibinfo {year} {2013})},\ \Eprint {http://arxiv.org/abs/1305.6610}
  {arXiv:1305.6610 [hep-ph]} \BibitemShut {NoStop}%
\bibitem [{\citenamefont {Turner}\ and\ \citenamefont
  {Wilczek}(1990)}]{Turner:1990rc}%
  \BibitemOpen
  \bibfield  {author} {\bibinfo {author} {\bibfnamefont {M.~S.}\ \bibnamefont
  {Turner}}\ and\ \bibinfo {author} {\bibfnamefont {F.}~\bibnamefont
  {Wilczek}},\ }\href {\doibase 10.1103/PhysRevLett.65.3080} {\bibfield
  {journal} {\bibinfo  {journal} {Phys.Rev.Lett.}\ }\textbf {\bibinfo {volume}
  {65}},\ \bibinfo {pages} {3080} (\bibinfo {year} {1990})}\BibitemShut
  {NoStop}%
\bibitem [{\citenamefont {Kosowsky}\ \emph
  {et~al.}(1992{\natexlab{a}})\citenamefont {Kosowsky}, \citenamefont
  {Turner},\ and\ \citenamefont {Watkins}}]{Kosowsky:1991ua}%
  \BibitemOpen
  \bibfield  {author} {\bibinfo {author} {\bibfnamefont {A.}~\bibnamefont
  {Kosowsky}}, \bibinfo {author} {\bibfnamefont {M.~S.}\ \bibnamefont
  {Turner}}, \ and\ \bibinfo {author} {\bibfnamefont {R.}~\bibnamefont
  {Watkins}},\ }\href {\doibase 10.1103/PhysRevD.45.4514} {\bibfield  {journal}
  {\bibinfo  {journal} {Phys.Rev.}\ }\textbf {\bibinfo {volume} {D45}},\
  \bibinfo {pages} {4514} (\bibinfo {year} {1992}{\natexlab{a}})}\BibitemShut
  {NoStop}%
\bibitem [{\citenamefont {Kosowsky}\ \emph
  {et~al.}(1992{\natexlab{b}})\citenamefont {Kosowsky}, \citenamefont
  {Turner},\ and\ \citenamefont {Watkins}}]{Kosowsky:1992rz}%
  \BibitemOpen
  \bibfield  {author} {\bibinfo {author} {\bibfnamefont {A.}~\bibnamefont
  {Kosowsky}}, \bibinfo {author} {\bibfnamefont {M.~S.}\ \bibnamefont
  {Turner}}, \ and\ \bibinfo {author} {\bibfnamefont {R.}~\bibnamefont
  {Watkins}},\ }\href {\doibase 10.1103/PhysRevLett.69.2026} {\bibfield
  {journal} {\bibinfo  {journal} {Phys.Rev.Lett.}\ }\textbf {\bibinfo {volume}
  {69}},\ \bibinfo {pages} {2026} (\bibinfo {year}
  {1992}{\natexlab{b}})}\BibitemShut {NoStop}%
\bibitem [{\citenamefont {Kosowsky}\ and\ \citenamefont
  {Turner}(1993)}]{Kosowsky:1992vn}%
  \BibitemOpen
  \bibfield  {author} {\bibinfo {author} {\bibfnamefont {A.}~\bibnamefont
  {Kosowsky}}\ and\ \bibinfo {author} {\bibfnamefont {M.~S.}\ \bibnamefont
  {Turner}},\ }\href {\doibase 10.1103/PhysRevD.47.4372} {\bibfield  {journal}
  {\bibinfo  {journal} {Phys.Rev.}\ }\textbf {\bibinfo {volume} {D47}},\
  \bibinfo {pages} {4372} (\bibinfo {year} {1993})},\ \Eprint
  {http://arxiv.org/abs/astro-ph/9211004} {arXiv:astro-ph/9211004 [astro-ph]}
  \BibitemShut {NoStop}%
\bibitem [{\citenamefont {Chialva}(2011)}]{Chialva:2010jt}%
  \BibitemOpen
  \bibfield  {author} {\bibinfo {author} {\bibfnamefont {D.}~\bibnamefont
  {Chialva}},\ }\href {\doibase 10.1103/PhysRevD.83.023512} {\bibfield
  {journal} {\bibinfo  {journal} {Phys.Rev.}\ }\textbf {\bibinfo {volume}
  {D83}},\ \bibinfo {pages} {023512} (\bibinfo {year} {2011})},\ \Eprint
  {http://arxiv.org/abs/1004.2051} {arXiv:1004.2051 [astro-ph.CO]} \BibitemShut
  {NoStop}%
\bibitem [{\citenamefont {Hawking}\ \emph {et~al.}(1982)\citenamefont
  {Hawking}, \citenamefont {Moss},\ and\ \citenamefont
  {Stewart}}]{Hawking:1982ga}%
  \BibitemOpen
  \bibfield  {author} {\bibinfo {author} {\bibfnamefont {S.}~\bibnamefont
  {Hawking}}, \bibinfo {author} {\bibfnamefont {I.}~\bibnamefont {Moss}}, \
  and\ \bibinfo {author} {\bibfnamefont {J.}~\bibnamefont {Stewart}},\ }\href
  {\doibase 10.1103/PhysRevD.26.2681} {\bibfield  {journal} {\bibinfo
  {journal} {Phys.Rev.}\ }\textbf {\bibinfo {volume} {D26}},\ \bibinfo {pages}
  {2681} (\bibinfo {year} {1982})}\BibitemShut {NoStop}%
\bibitem [{\citenamefont {Huber}\ and\ \citenamefont
  {Konstandin}(2008)}]{Huber:2008hg}%
  \BibitemOpen
  \bibfield  {author} {\bibinfo {author} {\bibfnamefont {S.~J.}\ \bibnamefont
  {Huber}}\ and\ \bibinfo {author} {\bibfnamefont {T.}~\bibnamefont
  {Konstandin}},\ }\href {\doibase 10.1088/1475-7516/2008/09/022} {\bibfield
  {journal} {\bibinfo  {journal} {JCAP}\ }\textbf {\bibinfo {volume} {0809}},\
  \bibinfo {pages} {022} (\bibinfo {year} {2008})},\ \Eprint
  {http://arxiv.org/abs/0806.1828} {arXiv:0806.1828 [hep-ph]} \BibitemShut
  {NoStop}%
\bibitem [{\citenamefont {Child}\ and\ \citenamefont
  {Giblin}(2012)}]{Child:2012qg}%
  \BibitemOpen
  \bibfield  {author} {\bibinfo {author} {\bibfnamefont {H.~L.}\ \bibnamefont
  {Child}}\ and\ \bibinfo {author} {\bibfnamefont {J.~T.}\ \bibnamefont
  {Giblin}},\ }\href {\doibase 10.1088/1475-7516/2012/10/001} {\bibfield
  {journal} {\bibinfo  {journal} {JCAP}\ }\textbf {\bibinfo {volume} {1210}},\
  \bibinfo {pages} {001} (\bibinfo {year} {2012})},\ \Eprint
  {http://arxiv.org/abs/1207.6408} {arXiv:1207.6408 [astro-ph.CO]} \BibitemShut
  {NoStop}%
\bibitem [{\citenamefont {Bodeker}\ and\ \citenamefont
  {Moore}(2009)}]{Bodeker:2009qy}%
  \BibitemOpen
  \bibfield  {author} {\bibinfo {author} {\bibfnamefont {D.}~\bibnamefont
  {Bodeker}}\ and\ \bibinfo {author} {\bibfnamefont {G.~D.}\ \bibnamefont
  {Moore}},\ }\href {\doibase 10.1088/1475-7516/2009/05/009} {\bibfield
  {journal} {\bibinfo  {journal} {JCAP}\ }\textbf {\bibinfo {volume} {0905}},\
  \bibinfo {pages} {009} (\bibinfo {year} {2009})},\ \Eprint
  {http://arxiv.org/abs/0903.4099} {arXiv:0903.4099 [hep-ph]} \BibitemShut
  {NoStop}%
\bibitem [{\citenamefont {Kamionkowski}\ \emph {et~al.}(1994)\citenamefont
  {Kamionkowski}, \citenamefont {Kosowsky},\ and\ \citenamefont
  {Turner}}]{Kamionkowski:1993fg}%
  \BibitemOpen
  \bibfield  {author} {\bibinfo {author} {\bibfnamefont {M.}~\bibnamefont
  {Kamionkowski}}, \bibinfo {author} {\bibfnamefont {A.}~\bibnamefont
  {Kosowsky}}, \ and\ \bibinfo {author} {\bibfnamefont {M.~S.}\ \bibnamefont
  {Turner}},\ }\href {\doibase 10.1103/PhysRevD.49.2837} {\bibfield  {journal}
  {\bibinfo  {journal} {Phys.Rev.}\ }\textbf {\bibinfo {volume} {D49}},\
  \bibinfo {pages} {2837} (\bibinfo {year} {1994})},\ \Eprint
  {http://arxiv.org/abs/astro-ph/9310044} {arXiv:astro-ph/9310044 [astro-ph]}
  \BibitemShut {NoStop}%
\bibitem [{\citenamefont {Espinosa}\ \emph {et~al.}(2010)\citenamefont
  {Espinosa}, \citenamefont {Konstandin}, \citenamefont {No},\ and\
  \citenamefont {Servant}}]{Espinosa:2010hh}%
  \BibitemOpen
  \bibfield  {author} {\bibinfo {author} {\bibfnamefont {J.~R.}\ \bibnamefont
  {Espinosa}}, \bibinfo {author} {\bibfnamefont {T.}~\bibnamefont
  {Konstandin}}, \bibinfo {author} {\bibfnamefont {J.~M.}\ \bibnamefont {No}},
  \ and\ \bibinfo {author} {\bibfnamefont {G.}~\bibnamefont {Servant}},\ }\href
  {\doibase 10.1088/1475-7516/2010/06/028} {\bibfield  {journal} {\bibinfo
  {journal} {JCAP}\ }\textbf {\bibinfo {volume} {1006}},\ \bibinfo {pages}
  {028} (\bibinfo {year} {2010})},\ \Eprint {http://arxiv.org/abs/1004.4187}
  {arXiv:1004.4187 [hep-ph]} \BibitemShut {NoStop}%
\bibitem [{\citenamefont {Caprini}\ and\ \citenamefont
  {Durrer}(2006)}]{Caprini:2006jb}%
  \BibitemOpen
  \bibfield  {author} {\bibinfo {author} {\bibfnamefont {C.}~\bibnamefont
  {Caprini}}\ and\ \bibinfo {author} {\bibfnamefont {R.}~\bibnamefont
  {Durrer}},\ }\href {\doibase 10.1103/PhysRevD.74.063521} {\bibfield
  {journal} {\bibinfo  {journal} {Phys.Rev.}\ }\textbf {\bibinfo {volume}
  {D74}},\ \bibinfo {pages} {063521} (\bibinfo {year} {2006})},\ \Eprint
  {http://arxiv.org/abs/astro-ph/0603476} {arXiv:astro-ph/0603476 [astro-ph]}
  \BibitemShut {NoStop}%
\bibitem [{\citenamefont {Gogoberidze}\ \emph {et~al.}(2007)\citenamefont
  {Gogoberidze}, \citenamefont {Kahniashvili},\ and\ \citenamefont
  {Kosowsky}}]{Gogoberidze:2007an}%
  \BibitemOpen
  \bibfield  {author} {\bibinfo {author} {\bibfnamefont {G.}~\bibnamefont
  {Gogoberidze}}, \bibinfo {author} {\bibfnamefont {T.}~\bibnamefont
  {Kahniashvili}}, \ and\ \bibinfo {author} {\bibfnamefont {A.}~\bibnamefont
  {Kosowsky}},\ }\href {\doibase 10.1103/PhysRevD.76.083002} {\bibfield
  {journal} {\bibinfo  {journal} {Phys.Rev.}\ }\textbf {\bibinfo {volume}
  {D76}},\ \bibinfo {pages} {083002} (\bibinfo {year} {2007})},\ \Eprint
  {http://arxiv.org/abs/0705.1733} {arXiv:0705.1733 [astro-ph]} \BibitemShut
  {NoStop}%
\bibitem [{\citenamefont {Caprini}\ \emph
  {et~al.}(2009{\natexlab{a}})\citenamefont {Caprini}, \citenamefont {Durrer},\
  and\ \citenamefont {Servant}}]{Caprini:2009yp}%
  \BibitemOpen
  \bibfield  {author} {\bibinfo {author} {\bibfnamefont {C.}~\bibnamefont
  {Caprini}}, \bibinfo {author} {\bibfnamefont {R.}~\bibnamefont {Durrer}}, \
  and\ \bibinfo {author} {\bibfnamefont {G.}~\bibnamefont {Servant}},\ }\href
  {\doibase 10.1088/1475-7516/2009/12/024} {\bibfield  {journal} {\bibinfo
  {journal} {JCAP}\ }\textbf {\bibinfo {volume} {0912}},\ \bibinfo {pages}
  {024} (\bibinfo {year} {2009}{\natexlab{a}})},\ \Eprint
  {http://arxiv.org/abs/0909.0622} {arXiv:0909.0622 [astro-ph.CO]} \BibitemShut
  {NoStop}%
\bibitem [{\citenamefont {Dolgov}\ \emph {et~al.}(2002)\citenamefont {Dolgov},
  \citenamefont {Grasso},\ and\ \citenamefont {Nicolis}}]{Dolgov:2002ra}%
  \BibitemOpen
  \bibfield  {author} {\bibinfo {author} {\bibfnamefont {A.~D.}\ \bibnamefont
  {Dolgov}}, \bibinfo {author} {\bibfnamefont {D.}~\bibnamefont {Grasso}}, \
  and\ \bibinfo {author} {\bibfnamefont {A.}~\bibnamefont {Nicolis}},\ }\href
  {\doibase 10.1103/PhysRevD.66.103505} {\bibfield  {journal} {\bibinfo
  {journal} {Phys.Rev.}\ }\textbf {\bibinfo {volume} {D66}},\ \bibinfo {pages}
  {103505} (\bibinfo {year} {2002})},\ \Eprint
  {http://arxiv.org/abs/astro-ph/0206461} {arXiv:astro-ph/0206461 [astro-ph]}
  \BibitemShut {NoStop}%
\bibitem [{\citenamefont {Hindmarsh}\ \emph {et~al.}(2014)\citenamefont
  {Hindmarsh}, \citenamefont {Huber}, \citenamefont {Rummukainen},\ and\
  \citenamefont {Weir}}]{Hindmarsh:2013xza}%
  \BibitemOpen
  \bibfield  {author} {\bibinfo {author} {\bibfnamefont {M.}~\bibnamefont
  {Hindmarsh}}, \bibinfo {author} {\bibfnamefont {S.~J.}\ \bibnamefont
  {Huber}}, \bibinfo {author} {\bibfnamefont {K.}~\bibnamefont {Rummukainen}},
  \ and\ \bibinfo {author} {\bibfnamefont {D.~J.}\ \bibnamefont {Weir}},\
  }\href {\doibase 10.1103/PhysRevLett.112.041301} {\bibfield  {journal}
  {\bibinfo  {journal} {Phys.Rev.Lett.}\ }\textbf {\bibinfo {volume} {112}},\
  \bibinfo {pages} {041301} (\bibinfo {year} {2014})},\ \Eprint
  {http://arxiv.org/abs/1304.2433} {arXiv:1304.2433 [hep-ph]} \BibitemShut
  {NoStop}%
\bibitem [{\citenamefont {Giblin}\ and\ \citenamefont
  {Mertens}(2014)}]{Giblin:2014qia}%
  \BibitemOpen
  \bibfield  {author} {\bibinfo {author} {\bibfnamefont {J.~T.}\ \bibnamefont
  {Giblin}}\ and\ \bibinfo {author} {\bibfnamefont {J.~B.}\ \bibnamefont
  {Mertens}},\ }\href {\doibase 10.1103/PhysRevD.90.023532} {\bibfield
  {journal} {\bibinfo  {journal} {Phys.Rev.}\ }\textbf {\bibinfo {volume}
  {D90}},\ \bibinfo {pages} {023532} (\bibinfo {year} {2014})},\ \Eprint
  {http://arxiv.org/abs/1405.4005} {arXiv:1405.4005 [astro-ph.CO]} \BibitemShut
  {NoStop}%
\bibitem [{\citenamefont {Kalaydzhyan}\ and\ \citenamefont
  {Shuryak}(2015)}]{Kalaydzhyan:2014wca}%
  \BibitemOpen
  \bibfield  {author} {\bibinfo {author} {\bibfnamefont {T.}~\bibnamefont
  {Kalaydzhyan}}\ and\ \bibinfo {author} {\bibfnamefont {E.}~\bibnamefont
  {Shuryak}},\ }\href {\doibase 10.1103/PhysRevD.91.083502} {\bibfield
  {journal} {\bibinfo  {journal} {Phys. Rev.}\ }\textbf {\bibinfo {volume}
  {D91}},\ \bibinfo {pages} {083502} (\bibinfo {year} {2015})},\ \Eprint
  {http://arxiv.org/abs/1412.5147} {arXiv:1412.5147 [hep-ph]} \BibitemShut
  {NoStop}%
\bibitem [{\citenamefont {Ghiglieri}\ and\ \citenamefont
  {Laine}(2015)}]{Ghiglieri:2015nfa}%
  \BibitemOpen
  \bibfield  {author} {\bibinfo {author} {\bibfnamefont {J.}~\bibnamefont
  {Ghiglieri}}\ and\ \bibinfo {author} {\bibfnamefont {M.}~\bibnamefont
  {Laine}},\ }\href {\doibase 10.1088/1475-7516/2015/07/022} {\bibfield
  {journal} {\bibinfo  {journal} {JCAP}\ }\textbf {\bibinfo {volume} {1507}},\
  \bibinfo {pages} {022} (\bibinfo {year} {2015})},\ \Eprint
  {http://arxiv.org/abs/1504.02569} {arXiv:1504.02569 [hep-ph]} \BibitemShut
  {NoStop}%
\bibitem [{\citenamefont {Ignatius}\ \emph {et~al.}(1994)\citenamefont
  {Ignatius}, \citenamefont {Kajantie}, \citenamefont {Kurki-Suonio},\ and\
  \citenamefont {Laine}}]{Ignatius:1993qn}%
  \BibitemOpen
  \bibfield  {author} {\bibinfo {author} {\bibfnamefont {J.}~\bibnamefont
  {Ignatius}}, \bibinfo {author} {\bibfnamefont {K.}~\bibnamefont {Kajantie}},
  \bibinfo {author} {\bibfnamefont {H.}~\bibnamefont {Kurki-Suonio}}, \ and\
  \bibinfo {author} {\bibfnamefont {M.}~\bibnamefont {Laine}},\ }\href
  {\doibase 10.1103/PhysRevD.49.3854} {\bibfield  {journal} {\bibinfo
  {journal} {Phys.Rev.}\ }\textbf {\bibinfo {volume} {D49}},\ \bibinfo {pages}
  {3854} (\bibinfo {year} {1994})},\ \Eprint
  {http://arxiv.org/abs/astro-ph/9309059} {arXiv:astro-ph/9309059 [astro-ph]}
  \BibitemShut {NoStop}%
\bibitem [{\citenamefont {Kurki-Suonio}\ and\ \citenamefont
  {Laine}(1996{\natexlab{a}})}]{KurkiSuonio:1995vy}%
  \BibitemOpen
  \bibfield  {author} {\bibinfo {author} {\bibfnamefont {H.}~\bibnamefont
  {Kurki-Suonio}}\ and\ \bibinfo {author} {\bibfnamefont {M.}~\bibnamefont
  {Laine}},\ }\href {\doibase 10.1103/PhysRevD.54.7163} {\bibfield  {journal}
  {\bibinfo  {journal} {Phys.Rev.}\ }\textbf {\bibinfo {volume} {D54}},\
  \bibinfo {pages} {7163} (\bibinfo {year} {1996}{\natexlab{a}})},\ \Eprint
  {http://arxiv.org/abs/hep-ph/9512202} {arXiv:hep-ph/9512202 [hep-ph]}
  \BibitemShut {NoStop}%
\bibitem [{\citenamefont {Enqvist}\ \emph {et~al.}(1992)\citenamefont
  {Enqvist}, \citenamefont {Ignatius}, \citenamefont {Kajantie},\ and\
  \citenamefont {Rummukainen}}]{Enqvist:1991xw}%
  \BibitemOpen
  \bibfield  {author} {\bibinfo {author} {\bibfnamefont {K.}~\bibnamefont
  {Enqvist}}, \bibinfo {author} {\bibfnamefont {J.}~\bibnamefont {Ignatius}},
  \bibinfo {author} {\bibfnamefont {K.}~\bibnamefont {Kajantie}}, \ and\
  \bibinfo {author} {\bibfnamefont {K.}~\bibnamefont {Rummukainen}},\ }\href
  {\doibase 10.1103/PhysRevD.45.3415} {\bibfield  {journal} {\bibinfo
  {journal} {Phys.Rev.}\ }\textbf {\bibinfo {volume} {D45}},\ \bibinfo {pages}
  {3415} (\bibinfo {year} {1992})}\BibitemShut {NoStop}%
\bibitem [{\citenamefont {Moore}\ and\ \citenamefont
  {Prokopec}(1995)}]{Moore:1995si}%
  \BibitemOpen
  \bibfield  {author} {\bibinfo {author} {\bibfnamefont {G.~D.}\ \bibnamefont
  {Moore}}\ and\ \bibinfo {author} {\bibfnamefont {T.}~\bibnamefont
  {Prokopec}},\ }\href {\doibase 10.1103/PhysRevD.52.7182} {\bibfield
  {journal} {\bibinfo  {journal} {Phys.Rev.}\ }\textbf {\bibinfo {volume}
  {D52}},\ \bibinfo {pages} {7182} (\bibinfo {year} {1995})},\ \Eprint
  {http://arxiv.org/abs/hep-ph/9506475} {arXiv:hep-ph/9506475 [hep-ph]}
  \BibitemShut {NoStop}%
\bibitem [{\citenamefont {Konstandin}\ \emph {et~al.}(2014)\citenamefont
  {Konstandin}, \citenamefont {Nardini},\ and\ \citenamefont
  {Rues}}]{Konstandin:2014zta}%
  \BibitemOpen
  \bibfield  {author} {\bibinfo {author} {\bibfnamefont {T.}~\bibnamefont
  {Konstandin}}, \bibinfo {author} {\bibfnamefont {G.}~\bibnamefont {Nardini}},
  \ and\ \bibinfo {author} {\bibfnamefont {I.}~\bibnamefont {Rues}},\ }\href
  {\doibase 10.1088/1475-7516/2014/09/028} {\bibfield  {journal} {\bibinfo
  {journal} {JCAP}\ }\textbf {\bibinfo {volume} {1409}},\ \bibinfo {pages}
  {028} (\bibinfo {year} {2014})},\ \Eprint {http://arxiv.org/abs/1407.3132}
  {arXiv:1407.3132 [hep-ph]} \BibitemShut {NoStop}%
\bibitem [{\citenamefont {Kurki-Suonio}\ and\ \citenamefont
  {Laine}(1996{\natexlab{b}})}]{KurkiSuonio:1996rk}%
  \BibitemOpen
  \bibfield  {author} {\bibinfo {author} {\bibfnamefont {H.}~\bibnamefont
  {Kurki-Suonio}}\ and\ \bibinfo {author} {\bibfnamefont {M.}~\bibnamefont
  {Laine}},\ }\href {\doibase 10.1103/PhysRevLett.77.3951} {\bibfield
  {journal} {\bibinfo  {journal} {Phys.Rev.Lett.}\ }\textbf {\bibinfo {volume}
  {77}},\ \bibinfo {pages} {3951} (\bibinfo {year} {1996}{\natexlab{b}})},\
  \Eprint {http://arxiv.org/abs/hep-ph/9607382} {arXiv:hep-ph/9607382 [hep-ph]}
  \BibitemShut {NoStop}%
\bibitem [{\citenamefont {Moore}\ and\ \citenamefont
  {Rummukainen}(2001)}]{Moore:2000jw}%
  \BibitemOpen
  \bibfield  {author} {\bibinfo {author} {\bibfnamefont {G.~D.}\ \bibnamefont
  {Moore}}\ and\ \bibinfo {author} {\bibfnamefont {K.}~\bibnamefont
  {Rummukainen}},\ }\href {\doibase 10.1103/PhysRevD.63.045002} {\bibfield
  {journal} {\bibinfo  {journal} {Phys.Rev.}\ }\textbf {\bibinfo {volume}
  {D63}},\ \bibinfo {pages} {045002} (\bibinfo {year} {2001})},\ \Eprint
  {http://arxiv.org/abs/hep-ph/0009132} {arXiv:hep-ph/0009132 [hep-ph]}
  \BibitemShut {NoStop}%
\bibitem [{\citenamefont {Linde}(1979)}]{Linde:1978px}%
  \BibitemOpen
  \bibfield  {author} {\bibinfo {author} {\bibfnamefont {A.~D.}\ \bibnamefont
  {Linde}},\ }\href {\doibase 10.1088/0034-4885/42/3/001} {\bibfield  {journal}
  {\bibinfo  {journal} {Rept.Prog.Phys.}\ }\textbf {\bibinfo {volume} {42}},\
  \bibinfo {pages} {389} (\bibinfo {year} {1979})}\BibitemShut {NoStop}%
\bibitem [{\citenamefont {Linde}(1983)}]{Linde:1981zj}%
  \BibitemOpen
  \bibfield  {author} {\bibinfo {author} {\bibfnamefont {A.~D.}\ \bibnamefont
  {Linde}},\ }\href {\doibase 10.1016/0550-3213(83)90293-6} {\bibfield
  {journal} {\bibinfo  {journal} {Nucl.Phys.}\ }\textbf {\bibinfo {volume}
  {B216}},\ \bibinfo {pages} {421} (\bibinfo {year} {1983})}\BibitemShut
  {NoStop}%
\bibitem [{\citenamefont {Hogan}(1983)}]{Hogan:1984hx}%
  \BibitemOpen
  \bibfield  {author} {\bibinfo {author} {\bibfnamefont {C.}~\bibnamefont
  {Hogan}},\ }\href {\doibase 10.1016/0370-2693(83)90553-1} {\bibfield
  {journal} {\bibinfo  {journal} {Phys.Lett.}\ }\textbf {\bibinfo {volume}
  {B133}},\ \bibinfo {pages} {172} (\bibinfo {year} {1983})}\BibitemShut
  {NoStop}%
\bibitem [{\citenamefont {Caprini}\ \emph
  {et~al.}(2009{\natexlab{b}})\citenamefont {Caprini}, \citenamefont {Durrer},
  \citenamefont {Konstandin},\ and\ \citenamefont {Servant}}]{Caprini:2009fx}%
  \BibitemOpen
  \bibfield  {author} {\bibinfo {author} {\bibfnamefont {C.}~\bibnamefont
  {Caprini}}, \bibinfo {author} {\bibfnamefont {R.}~\bibnamefont {Durrer}},
  \bibinfo {author} {\bibfnamefont {T.}~\bibnamefont {Konstandin}}, \ and\
  \bibinfo {author} {\bibfnamefont {G.}~\bibnamefont {Servant}},\ }\href
  {\doibase 10.1103/PhysRevD.79.083519} {\bibfield  {journal} {\bibinfo
  {journal} {Phys.Rev.}\ }\textbf {\bibinfo {volume} {D79}},\ \bibinfo {pages}
  {083519} (\bibinfo {year} {2009}{\natexlab{b}})},\ \Eprint
  {http://arxiv.org/abs/0901.1661} {arXiv:0901.1661 [astro-ph.CO]} \BibitemShut
  {NoStop}%
\bibitem [{\citenamefont {Figueroa}\ \emph {et~al.}(2013)\citenamefont
  {Figueroa}, \citenamefont {Hindmarsh},\ and\ \citenamefont
  {Urrestilla}}]{Figueroa:2012kw}%
  \BibitemOpen
  \bibfield  {author} {\bibinfo {author} {\bibfnamefont {D.~G.}\ \bibnamefont
  {Figueroa}}, \bibinfo {author} {\bibfnamefont {M.}~\bibnamefont {Hindmarsh}},
  \ and\ \bibinfo {author} {\bibfnamefont {J.}~\bibnamefont {Urrestilla}},\
  }\href@noop {} {\bibfield  {journal} {\bibinfo  {journal} {Phys. Rev. Lett.
  110,}\ }\textbf {\bibinfo {volume} {101302}} (\bibinfo {year} {2013})},\
  \Eprint {http://arxiv.org/abs/1212.5458} {arXiv:1212.5458 [astro-ph.CO]}
  \BibitemShut {NoStop}%
\bibitem [{\citenamefont {Arnold}\ \emph {et~al.}(2006)\citenamefont {Arnold},
  \citenamefont {Dogan},\ and\ \citenamefont {Moore}}]{Arnold:2006fz}%
  \BibitemOpen
  \bibfield  {author} {\bibinfo {author} {\bibfnamefont {P.~B.}\ \bibnamefont
  {Arnold}}, \bibinfo {author} {\bibfnamefont {C.}~\bibnamefont {Dogan}}, \
  and\ \bibinfo {author} {\bibfnamefont {G.~D.}\ \bibnamefont {Moore}},\ }\href
  {\doibase 10.1103/PhysRevD.74.085021} {\bibfield  {journal} {\bibinfo
  {journal} {Phys.Rev.}\ }\textbf {\bibinfo {volume} {D74}},\ \bibinfo {pages}
  {085021} (\bibinfo {year} {2006})},\ \Eprint
  {http://arxiv.org/abs/hep-ph/0608012} {arXiv:hep-ph/0608012 [hep-ph]}
  \BibitemShut {NoStop}%
\bibitem [{\citenamefont {Arnold}\ \emph {et~al.}(2000)\citenamefont {Arnold},
  \citenamefont {Moore},\ and\ \citenamefont {Yaffe}}]{Arnold:2000dr}%
  \BibitemOpen
  \bibfield  {author} {\bibinfo {author} {\bibfnamefont {P.~B.}\ \bibnamefont
  {Arnold}}, \bibinfo {author} {\bibfnamefont {G.~D.}\ \bibnamefont {Moore}}, \
  and\ \bibinfo {author} {\bibfnamefont {L.~G.}\ \bibnamefont {Yaffe}},\
  }\href@noop {} {\bibfield  {journal} {\bibinfo  {journal} {JHEP}\ }\textbf
  {\bibinfo {volume} {0011}},\ \bibinfo {pages} {001} (\bibinfo {year}
  {2000})},\ \Eprint {http://arxiv.org/abs/hep-ph/0010177}
  {arXiv:hep-ph/0010177 [hep-ph]} \BibitemShut {NoStop}%
\bibitem [{\citenamefont {Wilson}\ and\ \citenamefont
  {Matthews}(2003)}]{WilsonMatthews}%
  \BibitemOpen
  \bibfield  {author} {\bibinfo {author} {\bibfnamefont {J.}~\bibnamefont
  {Wilson}}\ and\ \bibinfo {author} {\bibfnamefont {G.}~\bibnamefont
  {Matthews}},\ }\href@noop {} {\emph {\bibinfo {title} {{Relativistic
  Numerical Hydrodyamics}}}}\ (\bibinfo  {publisher} {{Cambridge University
  Press}},\ \bibinfo {address} {Cambridge},\ \bibinfo {year}
  {2003})\BibitemShut {NoStop}%
\bibitem [{\citenamefont {Garcia-Bellido}\ \emph {et~al.}(2008)\citenamefont
  {Garcia-Bellido}, \citenamefont {Figueroa},\ and\ \citenamefont
  {Sastre}}]{GarciaBellido:2007af}%
  \BibitemOpen
  \bibfield  {author} {\bibinfo {author} {\bibfnamefont {J.}~\bibnamefont
  {Garcia-Bellido}}, \bibinfo {author} {\bibfnamefont {D.~G.}\ \bibnamefont
  {Figueroa}}, \ and\ \bibinfo {author} {\bibfnamefont {A.}~\bibnamefont
  {Sastre}},\ }\href {\doibase 10.1103/PhysRevD.77.043517} {\bibfield
  {journal} {\bibinfo  {journal} {Phys.Rev.}\ }\textbf {\bibinfo {volume}
  {D77}},\ \bibinfo {pages} {043517} (\bibinfo {year} {2008})},\ \Eprint
  {http://arxiv.org/abs/0707.0839} {arXiv:0707.0839 [hep-ph]} \BibitemShut
  {NoStop}%
\bibitem [{sup()}]{supplementary}%
  \BibitemOpen
  \href@noop {} {}\bibinfo {note} {See the supplementary material for plots of
  results from all the simulations discussed in this paper.}\BibitemShut
  {Stop}%
\bibitem [{\citenamefont {Caprini}\ \emph {et~al.}(2008)\citenamefont
  {Caprini}, \citenamefont {Durrer},\ and\ \citenamefont
  {Servant}}]{Caprini:2007xq}%
  \BibitemOpen
  \bibfield  {author} {\bibinfo {author} {\bibfnamefont {C.}~\bibnamefont
  {Caprini}}, \bibinfo {author} {\bibfnamefont {R.}~\bibnamefont {Durrer}}, \
  and\ \bibinfo {author} {\bibfnamefont {G.}~\bibnamefont {Servant}},\ }\href
  {\doibase 10.1103/PhysRevD.77.124015} {\bibfield  {journal} {\bibinfo
  {journal} {Phys.Rev.}\ }\textbf {\bibinfo {volume} {D77}},\ \bibinfo {pages}
  {124015} (\bibinfo {year} {2008})},\ \Eprint {http://arxiv.org/abs/0711.2593}
  {arXiv:0711.2593 [astro-ph]} \BibitemShut {NoStop}%
\bibitem [{\citenamefont {Kirzhnits}\ and\ \citenamefont
  {Linde}(1976)}]{Kirzhnits:1976ts}%
  \BibitemOpen
  \bibfield  {author} {\bibinfo {author} {\bibfnamefont {D.}~\bibnamefont
  {Kirzhnits}}\ and\ \bibinfo {author} {\bibfnamefont {A.~D.}\ \bibnamefont
  {Linde}},\ }\href {\doibase 10.1016/0003-4916(76)90279-7} {\bibfield
  {journal} {\bibinfo  {journal} {Annals Phys.}\ }\textbf {\bibinfo {volume}
  {101}},\ \bibinfo {pages} {195} (\bibinfo {year} {1976})}\BibitemShut
  {NoStop}%
\bibitem [{\citenamefont {Link}(1992)}]{Link:1992dm}%
  \BibitemOpen
  \bibfield  {author} {\bibinfo {author} {\bibfnamefont {B.}~\bibnamefont
  {Link}},\ }\href {\doibase 10.1103/PhysRevLett.68.2425} {\bibfield  {journal}
  {\bibinfo  {journal} {Phys.Rev.Lett.}\ }\textbf {\bibinfo {volume} {68}},\
  \bibinfo {pages} {2425} (\bibinfo {year} {1992})}\BibitemShut {NoStop}%
\bibitem [{\citenamefont {Huet}\ \emph {et~al.}(1993)\citenamefont {Huet},
  \citenamefont {Kajantie}, \citenamefont {Leigh}, \citenamefont {Liu},\ and\
  \citenamefont {McLerran}}]{Huet:1992ex}%
  \BibitemOpen
  \bibfield  {author} {\bibinfo {author} {\bibfnamefont {P.~Y.}\ \bibnamefont
  {Huet}}, \bibinfo {author} {\bibfnamefont {K.}~\bibnamefont {Kajantie}},
  \bibinfo {author} {\bibfnamefont {R.~G.}\ \bibnamefont {Leigh}}, \bibinfo
  {author} {\bibfnamefont {B.-H.}\ \bibnamefont {Liu}}, \ and\ \bibinfo
  {author} {\bibfnamefont {L.~D.}\ \bibnamefont {McLerran}},\ }\href {\doibase
  10.1103/PhysRevD.48.2477} {\bibfield  {journal} {\bibinfo  {journal}
  {Phys.Rev.}\ }\textbf {\bibinfo {volume} {D48}},\ \bibinfo {pages} {2477}
  (\bibinfo {year} {1993})},\ \Eprint {http://arxiv.org/abs/hep-ph/9212224}
  {arXiv:hep-ph/9212224 [hep-ph]} \BibitemShut {NoStop}%
\bibitem [{\citenamefont {Megevand}\ and\ \citenamefont
  {Membiela}(2014)}]{Megevand:2013yua}%
  \BibitemOpen
  \bibfield  {author} {\bibinfo {author} {\bibfnamefont {A.}~\bibnamefont
  {Megevand}}\ and\ \bibinfo {author} {\bibfnamefont {F.~A.}\ \bibnamefont
  {Membiela}},\ }\href {\doibase 10.1103/PhysRevD.89.103507} {\bibfield
  {journal} {\bibinfo  {journal} {Phys. Rev.}\ }\textbf {\bibinfo {volume}
  {D89}},\ \bibinfo {pages} {103507} (\bibinfo {year} {2014})},\ \Eprint
  {http://arxiv.org/abs/1311.2453} {arXiv:1311.2453 [astro-ph.CO]} \BibitemShut
  {NoStop}%
\bibitem [{\citenamefont {Brandenburg}\ \emph {et~al.}(1996)\citenamefont
  {Brandenburg}, \citenamefont {Enqvist},\ and\ \citenamefont
  {Olesen}}]{Brandenburg:1996fc}%
  \BibitemOpen
  \bibfield  {author} {\bibinfo {author} {\bibfnamefont {A.}~\bibnamefont
  {Brandenburg}}, \bibinfo {author} {\bibfnamefont {K.}~\bibnamefont
  {Enqvist}}, \ and\ \bibinfo {author} {\bibfnamefont {P.}~\bibnamefont
  {Olesen}},\ }\href {\doibase 10.1103/PhysRevD.54.1291} {\bibfield  {journal}
  {\bibinfo  {journal} {Phys. Rev.}\ }\textbf {\bibinfo {volume} {D54}},\
  \bibinfo {pages} {1291} (\bibinfo {year} {1996})},\ \Eprint
  {http://arxiv.org/abs/astro-ph/9602031} {arXiv:astro-ph/9602031 [astro-ph]}
  \BibitemShut {NoStop}%
\end{thebibliography}%

\end{document}